\documentclass[preprintnumbers,
a4paper,
superscriptaddress,
prd,nofootinbib,floatfix,12pt,
]{revtex4}

\usepackage{graphicx,amssymb,amsmath,color}
\usepackage{hyperref}
\usepackage{color}
\usepackage{setspace}
\usepackage{enumerate}

\usepackage[normalem]{ulem}  

\def\beq{\begin{equation}}
\def\eeq{\end{equation}}

\def\bea{\begin{eqnarray}}
\def\eea{\end{eqnarray}}
\def\bei{\begin{itemize}}
\def\eei{\end{itemize}}
\def\bmat{\begin{matrix}}
\def\emat{\end{matrix}}
\def\ble{\begin{flushleft}}
\def\ele{\end{flushleft}}
\def\={\,=\,}
\def\+{\,+\,}
\def\-{\,-\,}

\def\eg{{\it e.g.}}
\def\ie{{\it i.e.}}

\def\yhtt{\lambda_{h^0 {\tilde t_1} {\tilde t_1}^*}}

\def\order{{\cal{O}}}

\newcommand{\Fig}[1]{Fig.~\ref{#1}}
\newcommand{\Eq}[1]{Eq.~(\ref{#1})}
\newcommand{\Sec}[1]{Sec.~\ref{#1}}

\begin{document}

\title{Probing Light Stops with Stoponium}

\author{Brian Batell}
\email{brian.batell@cern.ch}
\affiliation{CERN, Theory Division, CH-1211 Geneva 23, Switzerland}

\author{Sunghoon Jung}
\email{nejsh21@gmail.com}
\affiliation{Korea Institute for Advanced Study, Seoul 130-722, Korea}

\begin{abstract} \vspace{3mm} \baselineskip=16pt
We derive new limits on light stops from diboson resonance searches in the $\gamma\gamma$, $Z \gamma$, $ZZ$, $WW$ and $hh$ channels from the first run of the LHC.
If the two-body decays of the light stop are mildly suppressed or kinematically forbidden, stoponium bound states will form in $pp$ collisions and subsequently decay via the pair annihilation of the constituent stops to diboson final states, yielding striking resonance signatures.
Remarkably, we find that stoponium searches are highly complementary to 
direct  collider searches and indirect probes of light stops such as Higgs coupling measurements.
Using an empirical quarkonia potential model and including the first two $S$-wave stoponium states, we find that in the decoupling limit $m_{\widetilde t_1} \lesssim 130$ GeV is excluded for \emph{any} value of the stop mixing angle and heavy stop mass by the combination of the latest resonance searches 
and the indirect constraints. 
The $\gamma \gamma$ searches are the most complementary to the indirect constraints, probing the stop ``blind spot'' parameter region in which the $h^0 \tilde t_1 \tilde t_1^*$ trilinear coupling is small. 
Interestingly, we also find that the $Z\gamma$ searches give a stronger constraint, $m_{\widetilde t_1} \lesssim 170$ GeV, if the stop is primarily left-handed. 
For a scenario with a bino LSP and stop NLSP, several gaps in the direct collider searches for stops can unambiguously be filled with the next run of the LHC.
For a stop LSP decaying through an R-parity violating $UDD$ coupling, the stoponium searches can fill the gap 100 GeV $\lesssim m_{\tilde t_1} \lesssim 200$ GeV in the direct searches for couplings $\lambda'' \lesssim 10^{-2}$.

\end{abstract}

\preprint{CERN-PH-TH-2015-070}
\preprint{KIAS-P15016}

\maketitle

\newpage

\baselineskip=15pt
\tableofcontents

\baselineskip=18pt

\section{Introduction}

Light stops are a quintessential feature of a natural supersymmetric theory~\cite{Dimopoulos:1995mi,Cohen:1996vb,Brust:2011tb,Papucci:2011wy}, being responsible for the cancellation of the dominant Higgs mass quadratic divergence coming from the top quark. However, after Run 1 of the LHC the possible existence of light stops has been strongly constrained by a suite of dedicated searches by ATLAS and CMS~\cite{Aad:2014qaa,ATLAS:stopsummary,CMS:2014yma,CMS:stopsummary}. For instance, in a simplified scenario containing a neutralino as the lightest supersymmetric particle (LSP) and the stop as the next-to-lightest supersymmetric particle (NLSP), with a few gaps and caveats, stop masses as high as $\sim 700$ GeV and LSP masses as high as $\sim 250$ GeV have been excluded~\cite{ATLAS:stopsummary,CMS:stopsummary}. Still, as is always the case with direct searches, the limits depend strongly on the spectrum and decays of the stops, and there remain open windows in which a light stop can hide from LHC searches. Examples include compressed~\cite{Martin:2008aw} or stealth~\cite{Fan:2011yu} stops and R-parity violating stop decays~\cite{Evans:2012bf,Bai:2013xla}. 
It is therefore critical to continue to develop new strategies to directly search for light stops~\cite{lightstops}. 

Another opportunity to probe light stops is presented when the stop is long lived, which naturally occurs in a number of motivated scenarios. Indeed, provided the stop does not have an \emph{unsuppressed} 2-body decay, stop pairs produced through gluon fusion can form a stoponium bound state due to the Coulombic attraction mediated by the strong force~\cite{stoponium,Drees:1993yr,Kats:2009bv,Kim:2014yaa,Martin:2008sv,Martin:2009dj,Younkin:2009zn,Kumar:2014bca}. 
Once the bound state is formed, it can decay via annihilation to $gg, \gamma\gamma, Z\gamma, WW, ZZ$ and $hh$, leaving a distinctive resonance signature. As we will emphasize below, the branching ratios to these final states depend essentially on one parameter, the trilinear Higgs-stop-stop ($h^0\tilde t_1 \tilde t_1^*$) coupling. Therefore, provided there is enough time for the stoponium to form and annihilate decay, the signature is independent of the details of the light SUSY spectrum and thus offers a highly complementary probe to direct stop searches. 

In this paper we will use the null results from $\gamma\gamma, Z\gamma, WW, ZZ$ and $hh$ resonance searches to constrain the production of stoponium during the first run of the LHC and in turn derive new limits on light stops. Assuming the stop is sufficiently long-lived, constraints on stoponium production in the $\gamma\gamma$, $\gamma Z$, and $ZZ$ channels {\it alone} limit stops lighter than $m_{\tilde t_1} \lesssim 125$ GeV for {\it any} value of the stop mixing angle and heavy stop mass $m_{\tilde t_2}$. Furthermore, for mostly left-handed light stops, the $Z\gamma$ searches constrain light stops up to about 170 GeV. We will also present projections for 14 TeV LHC with low (30 fb$^{-1}$) and high (3 ab$^{-1}$) luminosities.

One new aspect of our study which has not been emphasized in the literature is the interplay between stoponium searches and other indirect constraints on light stops. In particular, since the stoponium branching ratios depend primarily on the $h^0\tilde t_1 \tilde t_1^*$ coupling, our new constraints are highly complementary to those suggested by Higgs coupling measurements. Light stops with a large coupling to the Higgs boson significantly alter Higgs production in the gluon-gluon fusion channel and are already strongly constrained except for a ``blind spot'' in the parameter region where the $h^0\tilde t_1 \tilde t_1^*$ coupling is small~\cite{Craig:2014una,Fan:2014axa}. On the other hand, the $\gamma\gamma$ resonance signature of stoponium is enhanced in the region where the $h^0\tilde t_1 \tilde t_1^*$ coupling is small. Combining both constraints, we can rule out the existence of light stops up to about 130 GeV independently of the other parameters in the stop sector. Besides the interplay with Higgs couplings, we will also explore the impact of other indirect constraints on light stops, including vacuum stability and electroweak precision data. 

We also discuss the consequences of our results for two motivated scenarios in which stoponium annihilation decays are relevant. The first is the canonical R-parity conserving case with a bino LSP and a stop NLSP. For $m_{\tilde t_1} \lesssim m_{t}+ m_{\chi^0}$ the stoponium annihilation decays are visible, providing a clean and unambiguous probe of light stops which is independent of possible kinematic degeneracies or model dependencies of stop branching ratios. Second, we discuss the case of a stop LSP decaying via the R-parity violating $UDD$ operator to two jets, which is challenging to probe directly due to the large QCD backgrounds. Again, stoponium can provide a clear-cut test of light stops in this scenario. 

It is worth emphasizing that there is a potentially large source of theoretical uncertainty present in the stoponium production cross section coming from the assumed potential model and contribution of the excited states to the signal. The limits we present here are based on the empirical quarkonia potential model of Hagiwara et al.~\cite{Hagiwara:1990sq}. Furthermore, as discussed in Ref.~\cite{Younkin:2009zn}, it is likely that higher $S$-wave stoponium states contribute to the resonance signals, and we therefore include the first two states in our estimate. As we will argue below, we believe our assumptions in this regard are conservative, but we will also discuss in detail how different choices affect the stoponium production cross section and the resulting limits.

We begin in Section~\ref{sec:stoponium} with a brief review on the production and decay of stoponium at hadron colliders. Our new limits and future projections from LHC resonance searches on several benchmark scenarios for each diboson channel are presented in \Sec{sec:limits}.  The current limits from stoponium and other indirect constraints on the full stop sector parameter space are combined in \Sec{sec:interplay}. In \Sec{sec:scenarios} we discuss the implications of our results on two motivated scenarios containing lights stops. 
Section~\ref{sec:outlook} presents our conclusions and outlook. 
Finally we include several Appendices 
\ref{app:param}-\ref{app:searches} which
collect and summarize our conventions for the stop sector, the indirect constraints on light stops, the stop decay widths, our procedure for limit extrapolation, and the LHC diboson resonance searches used in our analysis.

\section{Stoponium Preliminaries}
\label{sec:stoponium}

In this work we will investigate the annihilation decay signatures of the $S$-wave $(J^{PC} = 0^{++})$ stoponium bound state, $\eta_{\tilde t}$, at the LHC. These signatures result from the following processes:
\begin{equation}
p p \rightarrow \eta_{\tilde t} + X, ~~ \eta_{\tilde t} \rightarrow \gamma\gamma, Z\gamma, WW, ZZ, h h \dots
\label{eq:stoponium}
\end{equation} 
Searches for stoponium resonances provide a highly complementary probe of light stops, testing different assumptions about SUSY spectra and parameters than both direct searches for stops and indirect probes such as Higgs coupling measurements, precision electroweak data, and vacuum stability constraints. In particular, direct searches for stops (and other new particles) are inherently model dependent, requiring assumptions about the superpartner spectrum and stop decay chains. 
In contrast, the limits and projections we will derive for stoponium will depend on two {\it different} assumptions: 1) the stop does not have an unsuppressed 2-body decay, and 2) the value of the trilinear Higgs-stop-stop coupling. As we will discuss shortly, the first assumption of a narrow stop width is actually satisfied in a number of interesting scenarios. The second assumption on the value of the $h^0 \tilde t_1 \tilde t_1$ coupling, which is determined by the stop sector masses and mixing angle, will govern the stoponium branching ratios for the various final states in Eq.~(\ref{eq:stoponium}). 

In this section we will collect the relevant ingredients entering into our study of stoponium.  
Our analysis relies on a number of previous works~\cite{stoponium,Drees:1993yr,Kats:2009bv,Kim:2014yaa}, notably those of Martin and collaborators~\cite{Martin:2008sv,Martin:2009dj,Younkin:2009zn,Kumar:2014bca}. Next we describe in more detail the basic requirements for observable stoponium annihilation decays. Following this we will review the production of stoponium at the LHC and the stoponium decay branching ratios. 

Before beginning, let us note here that we will work exclusively in this paper with the physical stop parameters $\{m_{\tilde t_1}, m_{\tilde t_2}, \theta_t\}$, in which $\tilde t_1$ $(\tilde t_2)$ denotes the lightest (heaviest) physical stop state, and a mixing angle $\theta = 0\, (\pi/2)$ corresponds to a purely left-handed (right-handed) stop. We will also restrict to the decoupling limit. We refer the reader to Appendix~\ref{app:param} for our stop sector conventions.

\subsection{Conditions for stoponium formation and annihilation decays}

What conditions are required for the distinctive resonance signatures of stoponium to be observable?\,\footnote{Useful related discussions for toponium bound states can be found in Refs.~\cite{Fadin:1987wz,Kats:2009bv}.} Provided they are kinematically accessible, stop pairs will be copiously produced in $pp$ collisions. However, whether or not the bound state forms and subsequently decays via the annihilation processes in Eq.~(\ref{eq:stoponium}) depends on the natural width of the stop. The bound state formation time is on the order of the inverse binding energy which, assuming a Coulombic potential, is given by $E_b  \sim  \alpha_s(a_0^{-1})^2 \, m_{\tilde t_1} \sim \order(1\,{\rm  GeV} )$, where $a_0 \sim [\alpha_s(a_0^{-1}) m_{\tilde t}]^{-1}$ is the ``Bohr radius'' of the system. Therefore, stoponium will form provided the natural width of the stop is smaller than the binding energy. Subsequently, the decay of the bound state may proceed either through the prompt decay of the constituent stop or through the annihilation decays. 
For instance, if the $h^0\tilde t_1 \tilde t_1^*$ coupling is small the dominant annihilation decay is to a pair of gluons, with partial width  (see also \Sec{sec:branching} below)
\begin{equation}
\Gamma(\eta_{\tilde t}\rightarrow gg) \simeq  \frac{4}{3} \alpha_s^2 \frac{|R(0)|^2}{m^2_{\eta_{\tilde t}}},
\label{eq:gg}
\end{equation}
where $R(0) = \sqrt{4 \pi} \psi(0)$ is the radial wavefunction at the origin, for which we employ the parameterization in Ref.~\cite{Hagiwara:1990sq,Martin:2008sv}. For $m_{\tilde t_1} \sim \order(100)$ GeV, the factor  $|R(0)|^2/m^2_{\eta_{\tilde t}} \sim 0.2$ GeV, leading to a partial width $\Gamma(\eta_{\tilde t}\rightarrow gg) \sim \order(1)$ MeV in the light stop mass range of interest. 
If the annihilation decay width dominates over the natural stop width we can potentially observe the stoponium through its resonance signature. 

In summary we require,
\begin{equation}
\Gamma_{\widetilde t_1} \lesssim \Gamma_{\eta_{\widetilde t}},  E_{b}.
\label{eq:cond}
\end{equation}
In practice, since the annihilation decay width is typically smaller than the binding energy, it sets the upper bound in Eq.~(\ref{eq:cond}). 
In fact, these conditions are satisfied in a number of interesting of scenarios and over a wide range of model parameters, particularly when the stops are light. We will examine two such scenarios in Section~\ref{sec:scenarios}.

\subsection{Production}
\label{sec:prod}

At leading order (LO) the production of stoponium in $pp$ collisions at the LHC proceeds via gluon-gluon fusion, with a cross section given by~\cite{Martin:2008sv}
\begin{equation}
\sigma_{\rm LO}(pp \rightarrow \eta_{\tilde t}) = \frac{\pi^2}{8 m_{ \eta_{\tilde t} }^3  } \Gamma(\eta_{\tilde t} \rightarrow gg) 
{\cal P}_{gg}(\tau),
\label{eq:LO}
\end{equation}
where 
${\cal P}_{gg}$ is the gluon parton luminosity as a function of $\tau \equiv m_{\eta_{\tilde t}}/s$, with $s$ the squared center-of-mass energy, defined in Eq.~(\ref{eq:Pab}) in Appendix \ref{app:extrap}.
The partial decay width for $\Gamma(\eta_{\tilde t} \rightarrow gg)$ is given in Eq.~(\ref{eq:gg}). Next-to-leading order (NLO) perturbative QCD corrections to stoponium production have been computed in 
Ref.~\cite{Younkin:2009zn} and with the threshold resummation in Ref.~\cite{Kim:2014yaa}. To account for these corrections we will use the results of Ref.~\cite{Younkin:2009zn}. In particular, we extract the cross section prediction from their Fig.\,9 and furthermore estimate a conservative $\sim 10 \%$ scale uncertainty on this prediction from their Fig.\,7~\cite{Younkin:2009zn}. We can also reasonably estimate a $\sim 5-10\%$ PDF uncertainty in the mass range of interest, similarly to that for Higgs production via gluon-gluon fusion~\cite{HiggsXS}.

There are likely larger sources of theoretical uncertainties than the QCD scale uncertainty, which are, however, difficult to quantify. Notice in particular that the cross section in Eq.~(\ref{eq:LO}) depends on the value of the stoponium wavefunction at the origin squared, $|R(0)|^2$, through Eq.~(\ref{eq:gg}), and our imprecise understanding of this matrix element potentially provides a significant source of uncertainty. 
Ref.~\cite{Younkin:2009zn} employs the $\Lambda_{\overline{\rm MS}}^{(4)} = 300$ MeV parameterization of the wavefunction at the origin from the study of Hagiwara et al.~\cite{Hagiwara:1990sq}. This parameterization is based on an empirical quarkonia potential model, which deviates significantly from the pure Coulombic form at large $m_{\tilde t_1}$ due to the effect of higher order QCD corrections. For instance, the prediction for $|R(0)|^2$ for the ground state is smaller by roughly a factor of 2 in comparison to that obtained from a Coulomb potential, suggesting that our estimates of the production rate are conservative. Ref.~\cite{Hagiwara:1990sq} also present $\Lambda_{\overline{\rm MS}}^{(4)} = 200$ MeV and $\Lambda_{\overline{\rm MS}}^{(4)} = 400$ MeV parameterizations, which lead to a $\sim 10\%$ reduction or enhancement in the stoponium production rate, respectively, giving some sense of the uncertainties involved. We note that the study of Ref.~\cite{Hagiwara:1990sq} is 25 years old, and given the significant experimental and theoretical progress in the studies of quarkonia and QCD in the interim, it would be extremely useful to perform a modern analysis in the spirit of~\cite{Hagiwara:1990sq}, with the particular aim of assessing the theoretical uncertainties.  

\begin{figure}[t] \centering 
\includegraphics[width=0.49\textwidth]{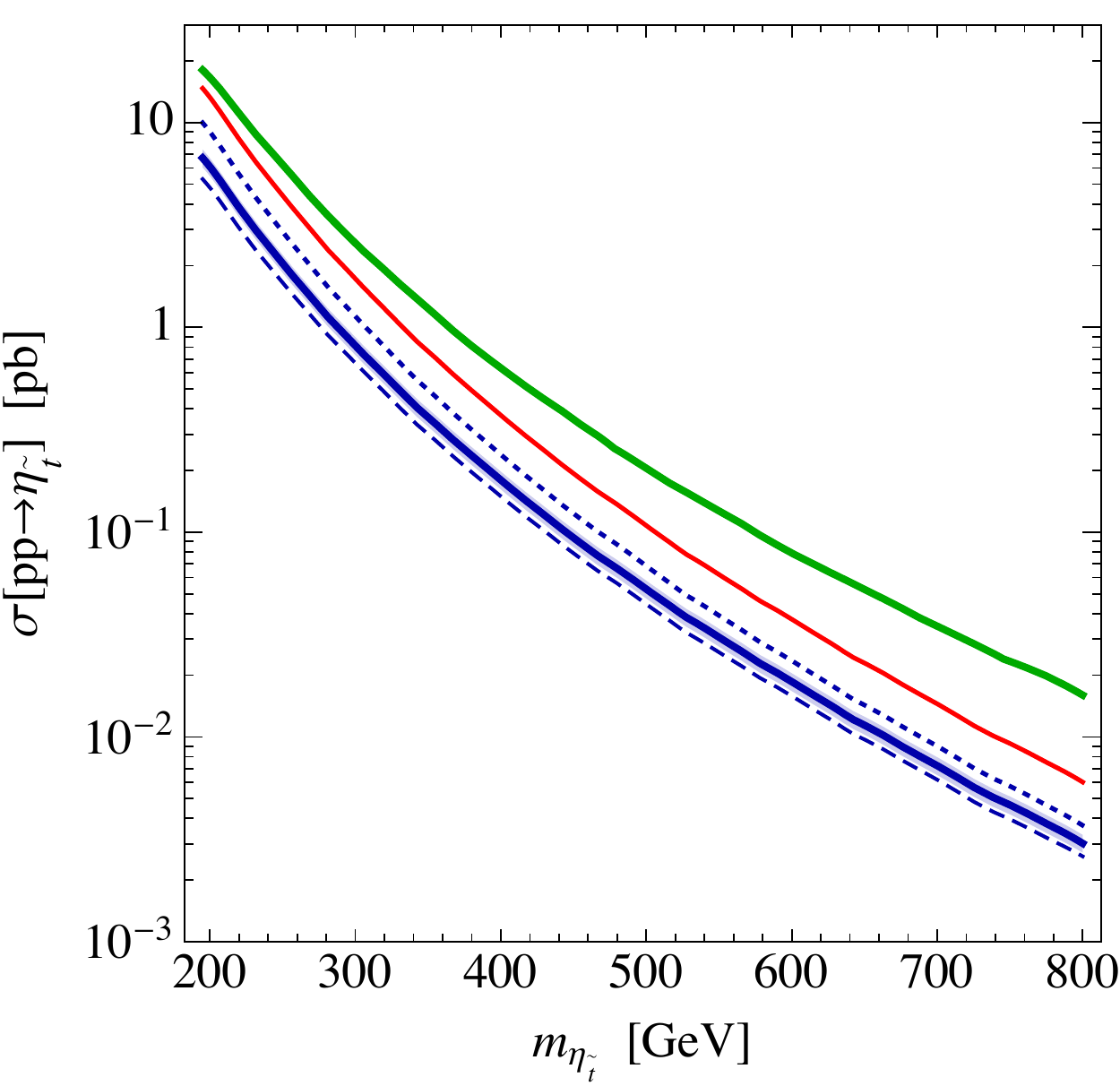}
\caption{\baselineskip=16pt 
Stoponium production cross sections at the LHC. 
Our baseline cross section for $\sqrt{s} = 8$ TeV (solid blue) and $\sqrt{s} = 14$ TeV (solid green) from Ref.~\cite{Younkin:2009zn}, which assume the 
$\Lambda_{\overline{\rm MS}}^{(4)} = 300$ MeV parameterization of the stoponium wavefunction at the origin from Ref.~\cite{Hagiwara:1990sq} and include the contribution of the first two $S$-wave states. These cross sections are used to derive our limits and projections.  
For $\sqrt{s}=8$ TeV we also display for comparison the cross section derived with the quarkonium potential~\cite{Hagiwara:1990sq} including instead only the ground state (blue dashed) and the first 10 $S$-wave states (blue dotted), as well as the cross section assuming a Coulomb potential (red) (including only the ground state) with $\alpha_S$ evaluated at
$\mu = 1/\langle r_{1S} \rangle = 2 \, \alpha_S(\mu) \, m_{\eta_{\tilde t}}/9$ .
}
\label{fig:XS}
\end{figure}

Another source of uncertainty in the prediction is the contribution of higher $S$-wave stoponium states to the annihilation decay signals. Ref.~\cite{Younkin:2009zn} includes the first two states in their calculation, but reasonably argues that higher states will likely add to the signal, and in this way our estimates of the production rate are likely conservative. 
In Figure~\ref{fig:XS} we display the stoponium production cross section at the LHC under several different assumptions regarding the potential and contribution of excited states.  
Needless to say, it would be extremely interesting to study the stoponium system in more detail to both quantify the uncertainty due to the modelling of the bound state potential as well as understand the phenomenological consequences of the higher radial states.

\subsection{Decay branching ratios} 
\label{sec:branching}

For the stoponium decay branching ratios, we use the LO results presented in Ref.~\cite{Martin:2008sv} for all decay channels except for the $\eta_{\tilde t} \rightarrow \gamma \gamma$ channel. For the diphoton channel we obtain the partial width $\Gamma(\eta_{\tilde t} \rightarrow \gamma \gamma)$ by multiplying the LO partial width $\Gamma_{\rm LO}(\eta_{\tilde t} \rightarrow gg)$~\cite{Martin:2008sv} by the NLO
ratio $\Gamma_{\rm NLO}(\eta_{\tilde t} \rightarrow \gamma \gamma)/\Gamma_{\rm NLO}(\eta_{\tilde t} \rightarrow gg)$ computed in Ref.~\cite{Martin:2009dj}. This approximation reproduces the correct NLO BR($\gamma \gamma$) when BR($gg$) dominates, and it is this case in which the diphoton channel is most important\footnote{Although the $Z\gamma$ channel is also important in this parameter region, the NLO corrections for this channel are not available.}. When other decay modes such as $hh$ or $WW$ dominate instead, the diphoton branching ratio is anyway too small to induce an observable signal. We use the two-loop running $\alpha_S(\mu)$ and the one-loop running $\alpha(\mu)$ renormalized from $\alpha_S(m_Z)=0.118$ and $\alpha(m_Z) = 1/128.0$.

\begin{figure}[t] \centering 
\includegraphics[width=0.49\textwidth]{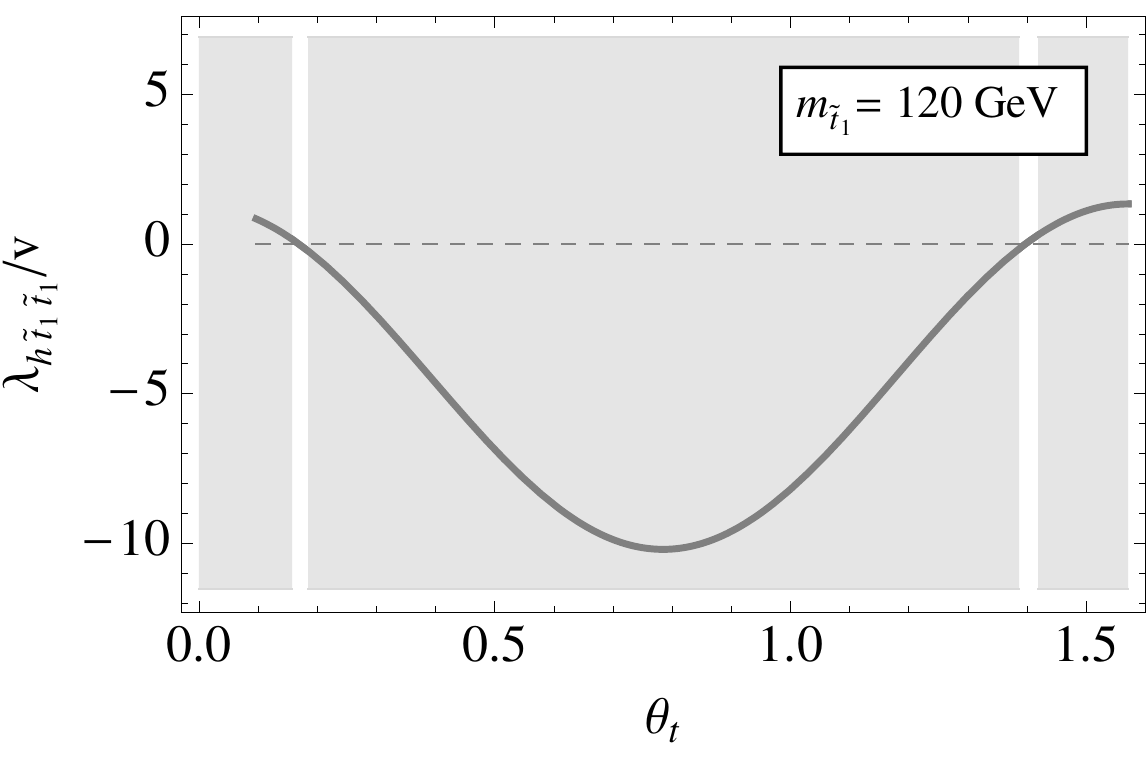} 
\includegraphics[width=0.49\textwidth]{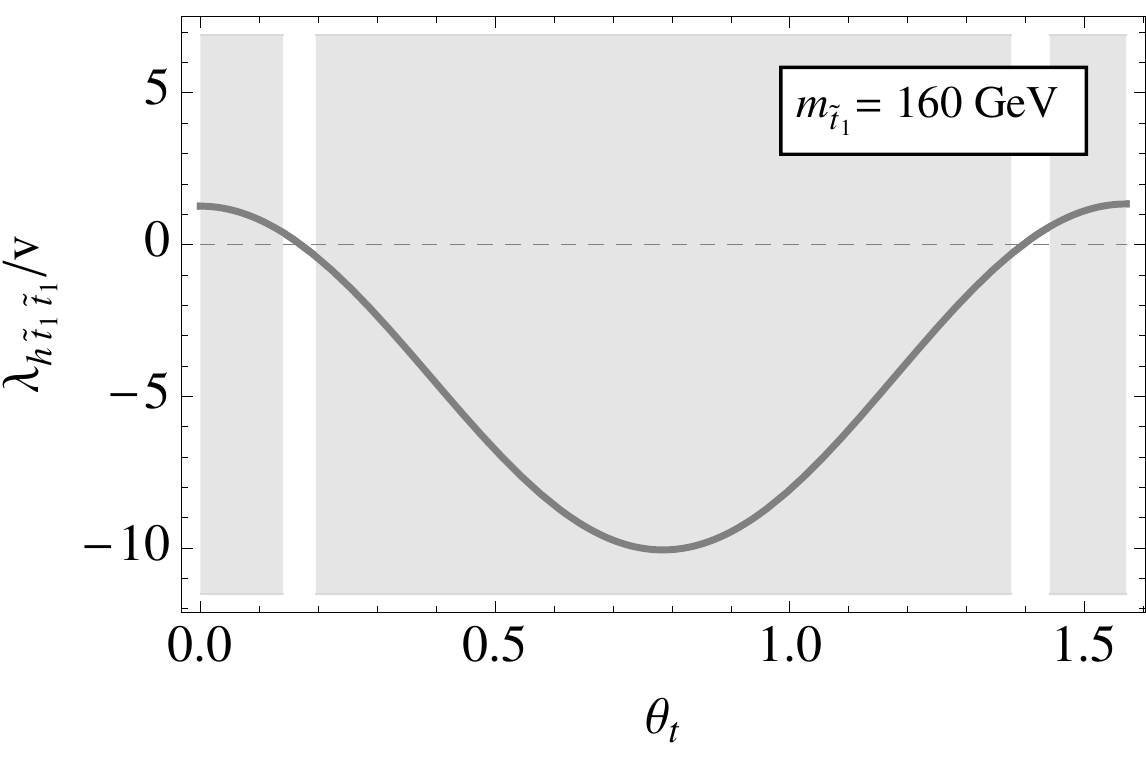}\\
\includegraphics[width=0.49\textwidth]{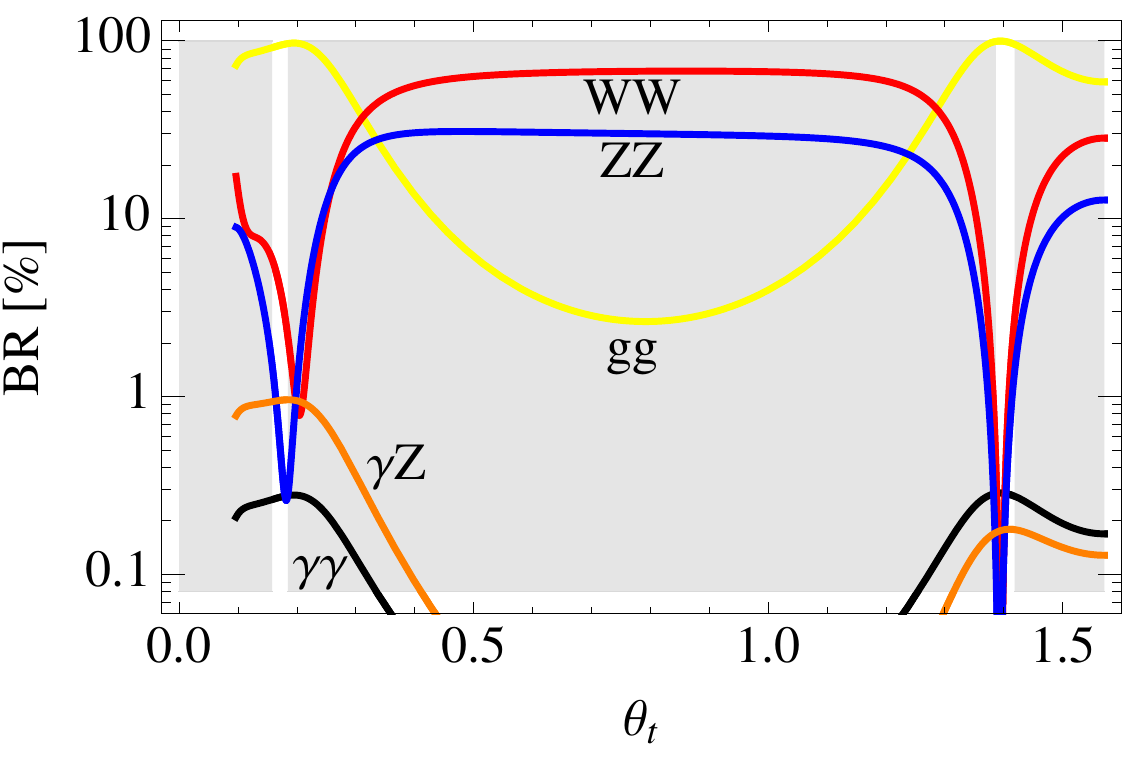} 
\includegraphics[width=0.49\textwidth]{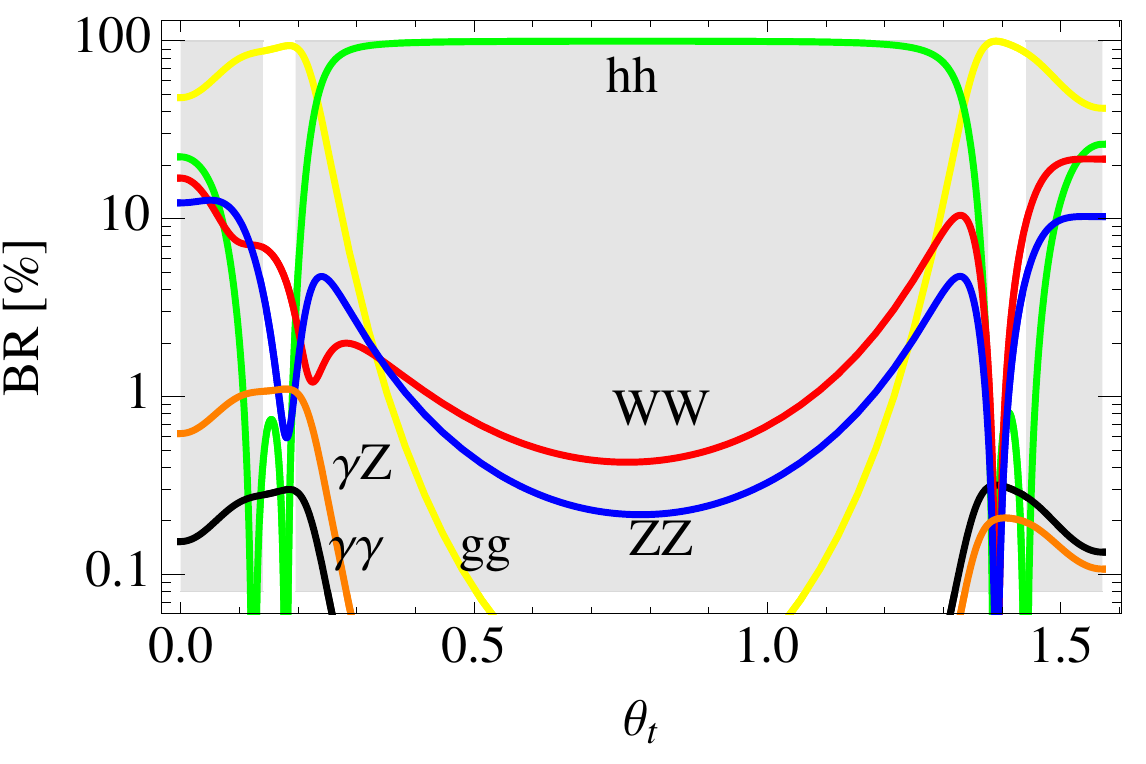}
\caption{\baselineskip=16pt 
The coupling $\yhtt/v$ (upper) and stoponium branching ratios (lower) for $m_{\widetilde t_1}=$ 120 GeV (left) and 160 GeV (right) as functions of the stop mixing angle. A strong correlation between $\yhtt$ and the stoponium branching ratios is observed. The shaded regions are constrained indirectly by Higgs signal strength data, precision electroweak measurements, and vacuum stability  as discussed in Appendix \ref{sec:indirect}. Here we have fixed $m_{\widetilde t_2}=1$ TeV. Note that in the left panel we have terminated the plot at small $\theta_t$ where the sbottom becomes tachyonic.}
\label{fig:BR}
\end{figure}

We will work in the decoupling limit. The minimal low energy spectrum will contain the stops and possibly the left-handed sbottom. It is also possible that the spectrum contains a bino-like neutralino LSP, since the condition (\ref{eq:cond}) is readily satisfied when $m_{\tilde t_1} \lesssim m_t + m_{\chi^0}$  (see \Sec{sec:scenarios} for a detailed discussion of this scenario), although the stoponium annihilation decays to neutralinos in this case are always negligible. We will not consider Higgsinos or Winos that are lighter than $\tilde t_1$, as in this case we expect the stop to have an unsuppressed two body decay to a lighter chargino and a bottom quark, generically violating the condition 
(\ref{eq:cond})\footnote{In any case, such scenarios with the Higgsino or Wino lighter than the stop are already strongly constrained by searches of $b\bar{b}$ plus missing energy~\cite{Aad:2013ija}.}
All other heavy Higgs scalars and superpartners are taken heavy and decoupled and will not influence the stoponium decays. The mass of the left-handed sbottom is fixed by assuming vanishing sbottom mixing (see \Eq{eq:sbottom}). 
We emphasize that left-handed sbottoms
can have non-negligible effects on the stoponium branching ratios when the lightest stop is primarily left-handed. As typically happens, unitarity and gauge invariance induce a (partial) cancellation in the $WW$ partial width between $t$-channel stop and sbottom exchanges if the stop is mostly left-handed~\cite{Drees:1993yr}. 
Finally, the stoponium decays to $b\bar{b}$ and $t\bar{t}$ are also subdominant and will not be considered further. 

The main decay modes, $\eta_{\tilde t}\rightarrow gg, \, \gamma \gamma,\, Z\gamma, \, WW, \, ZZ$ and $hh$, are described well by the three physical stop parameters, $ m_{\widetilde t_1}, \, m_{\widetilde t_2}, \, \theta_t$, and $t_\beta \equiv \tan \beta$. These parameters are defined in Appendix~\ref{app:param}. The dependence on $t_\beta$ is weak in most cases, so we fix $t_\beta=10$ throughout in this paper. 
The trilinear coupling of the 125 GeV Higgs to the light stops, $\yhtt$, plays a crucial role in determining the stoponium decay pattern. This coupling is written in terms of the physical stop parameters as
\bea
\yhtt &=& \sqrt{2}v \, \left \{ \frac{m_t^2}{v^2} \+ \frac{m_Z^2 c_{2\beta} }{v^2} \left[ c_t^2 \left(  \frac{1}{2} - \frac{2}{3} s_W^2 \right) + s_t^2 \left( \frac{2}{3} s_W^2 \right) \right]  \+ s_t^2 c_t^2 \frac{ m_{\tilde t_1}^2 - m_{\tilde t_2}^2 }{v^2} \right\}. \label{eq:lamhtt} 
\eea
Due to the last term in \Eq{eq:lamhtt}, this coupling
can become much larger than the electroweak scale, $v = 174$ GeV, when the stop mixing is large, inducing sizable partial decay widths for stoponium to $WW,\, ZZ,\, hh$ via the $s$-channel exchange of $h^0$. 
This is illustrated in \Fig{fig:BR}, where we display the mixing angle dependence of the coupling $\yhtt$ and the stoponium branching ratios. A strong correlation is observed between the value of $\yhtt$ and the pattern of branching ratios. For large values of mixing, the coupling $\yhtt$ grows and the stoponium dominantly decays to $WW$, $ZZ$, and, if kinematically allowed, $hh$. Instead, when $\yhtt$ is small, the $gg$ decay dominates and, importantly, the $\gamma\gamma$ and $Z\gamma$ branching ratios are maximized (see \Sec{sec:diphoton} and \Sec{sec:zgamma} for further discussion).

Large values of $\yhtt$ are constrained, particularly for light stops, by indirect tests including Higgs signal strength measurements, electroweak precision data, and vacuum stability. These indirect constraints are reviewed in Appendix \ref{sec:indirect} and shown as the shaded regions in \Fig{fig:BR}. In particular, we observe that only a small region around $\yhtt = 0$ is consistent with these indirect constraints for very light stops. 
It is therefore very interesting to consider the $\eta_{\tilde t}\rightarrow\gamma \gamma$ (and $Z\gamma$) channel as it can provide an independent probe of this unconstrained region of parameter space.

\section{New limits on light stops and future prospects}
\label{sec:limits}

We now derive the current limits on light stops from stoponium annihilation decays and estimate the future reach of each diboson channel in Eq.~(\ref{eq:stoponium}): $\gamma \gamma, \, Z\gamma, \, WW, \, ZZ$ and $hh$.  
In this section, we simply assume that the stoponium forms and decays via annihilation, but we will discuss particular models and relevant parameter space in \Sec{sec:scenarios}. For each diboson channel, we define and consider the {\it ideal} or {\it benchmark} branching ratio that best represents the optimal or characteristic reach, respectively, over a wide range of the stop masses. Based on these benchmarks, we 
present the limits as a function of the stoponium mass. 
Next, in \Sec{sec:interplay} we exhibit the exclusion limits in the general stop parameter space, \ie, as a function of the stop masses and mixing angles, accounting for the realistic branching ratios and comparing with the indirect constraints described in Appendix~\ref{sec:indirect}.

Our limits and projections are obtained by recasting the results of the latest heavy resonance searches at LHC 7+8 TeV, which are collected in Appendix~\ref{app:searches}. We will present the strongest limit in each diboson channel. For the future sensitivities, we extrapolate the current expected limits 
to LHC 14 TeV with 30 fb$^{-1}$ and 3 ab$^{-1}$.  Our extrapolation method and assumptions on statistical and systematic uncertainties are described in Appendix~\ref{app:extrap}. 

\subsection{$\gamma \gamma$} \label{sec:diphoton}

\begin{figure}[t] \centering 
\includegraphics[width=0.49\textwidth]{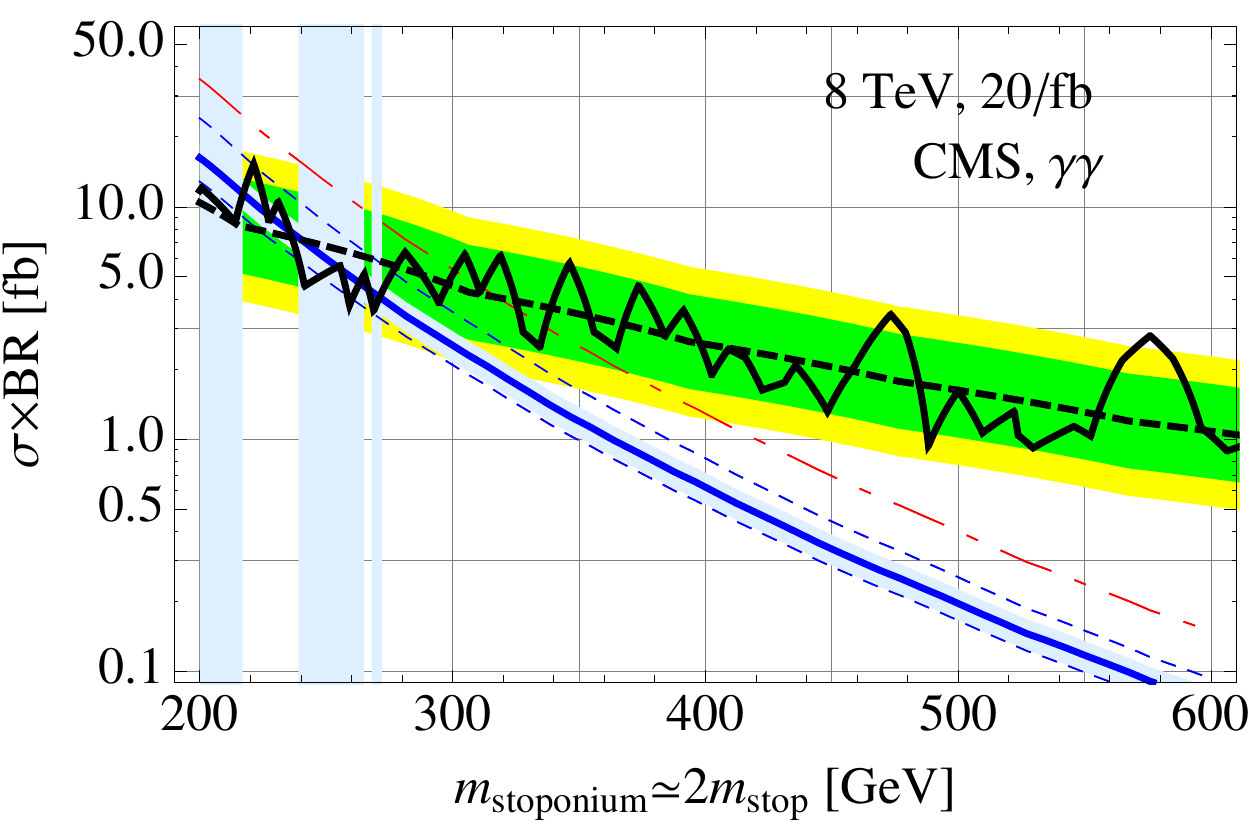}
\includegraphics[width=0.49\textwidth]{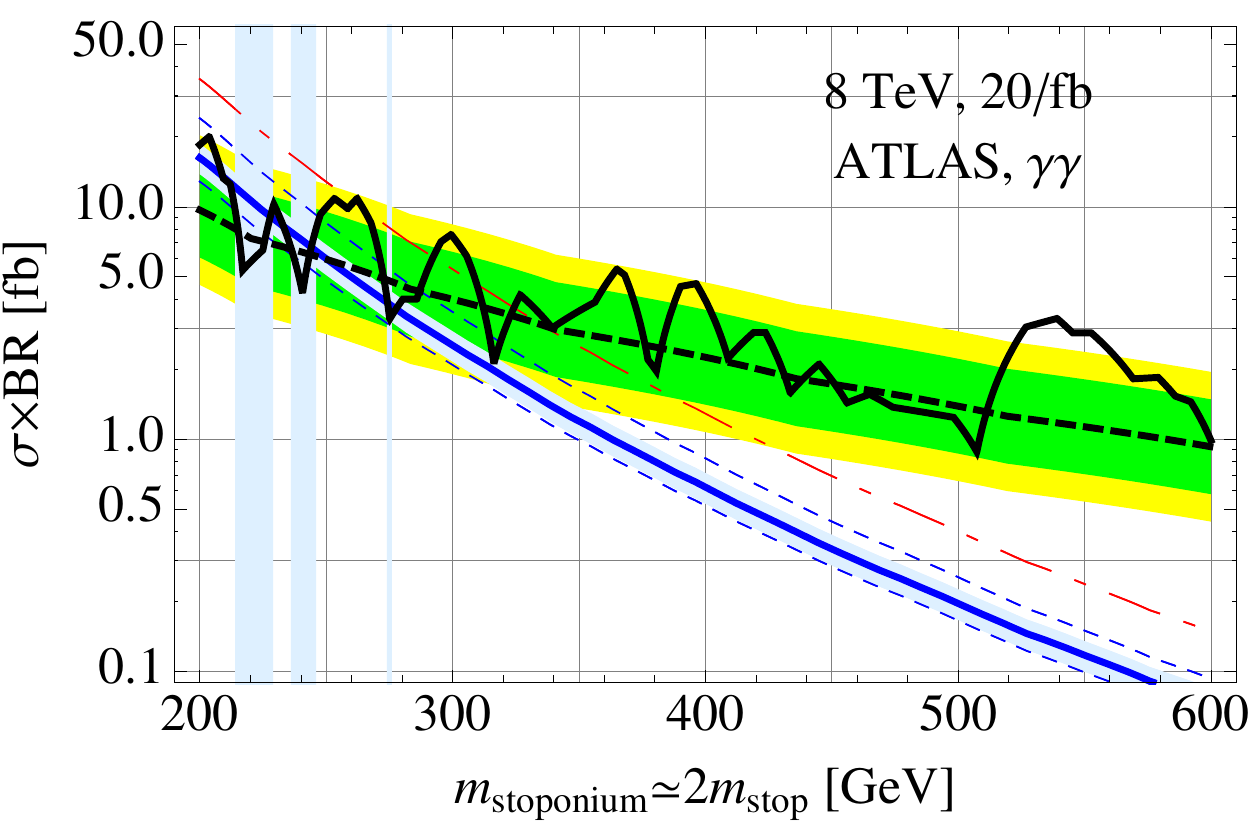}
\includegraphics[width=0.49\textwidth]{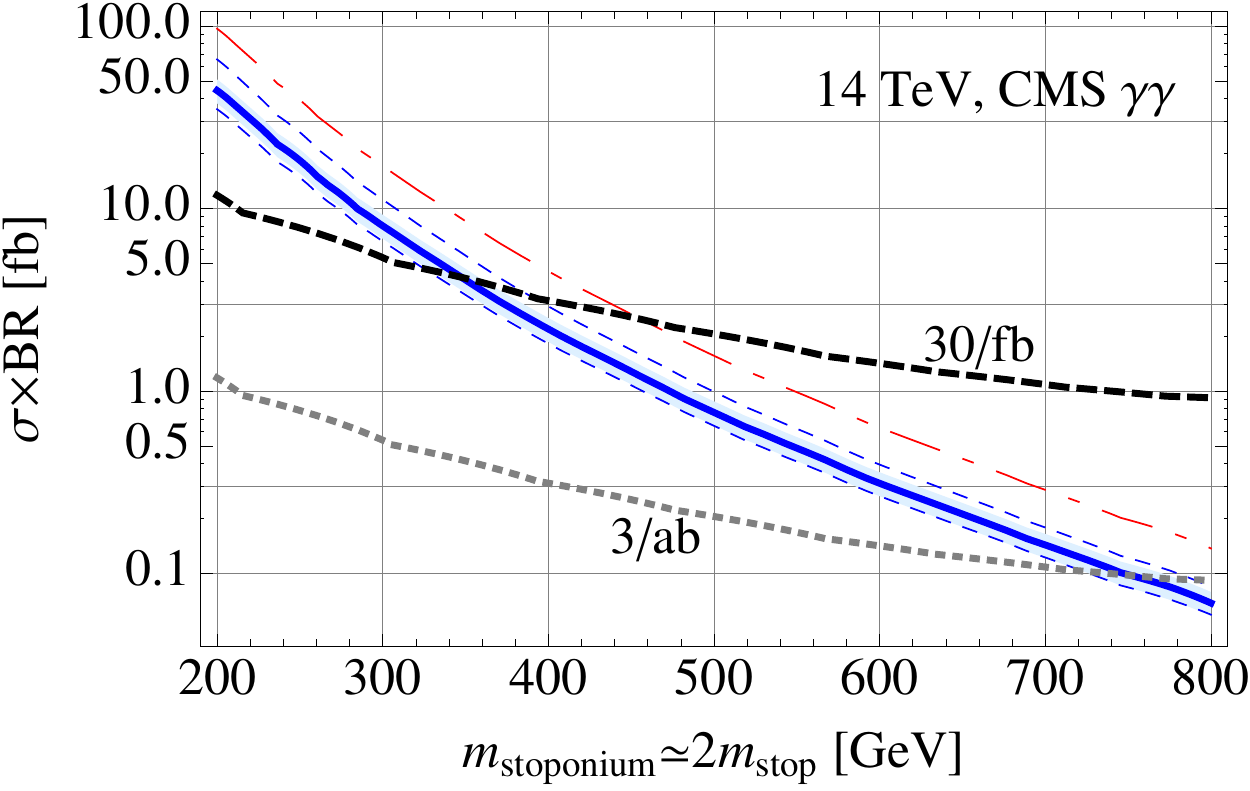}
\includegraphics[width=0.49\textwidth]{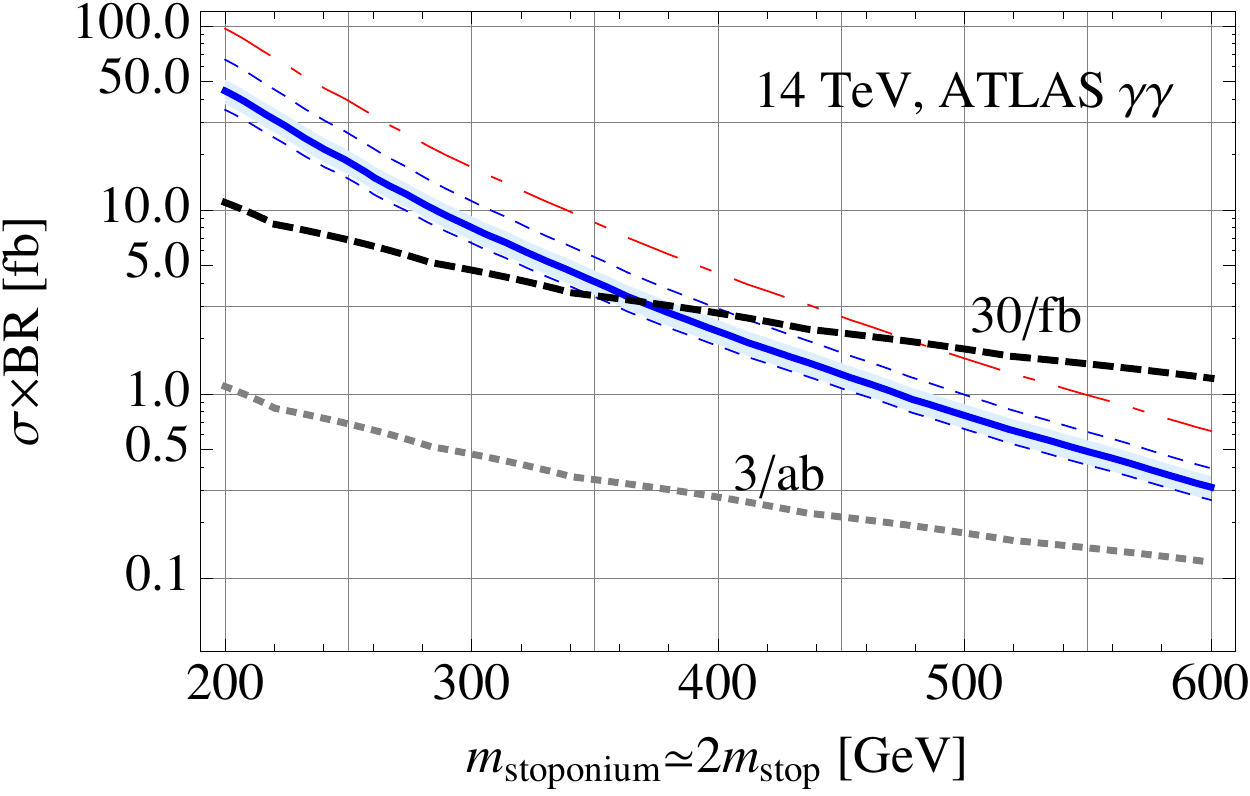}
\caption{\baselineskip=16pt 
Current bounds (upper) and future sensitivity (lower) on the ideal $\gamma \gamma$ scenario defined in \Sec{sec:diphoton}. ATLAS\,(right) and CMS\,(left) $\gamma \gamma$ resonance search results are used; the CMS result is extended to a higher mass. Four theoretical predictions are shown as in \Fig{fig:XS}: considering the quarkonium-inspired potential of Ref.~\cite{Hagiwara:1990sq} and including first two $S$-wave states (thick solid), 1S only (bottom dashed), first 10 $S$-wave states (mid-dashed) as well as considering a 1S with Coulomb potential (top dot-dashed). A $\pm 15\%$ scale+PDF uncertainty is also shown as a  band around the thick line. Vertical shaded regions are excluded if 1S+2S states are considered. A constant fiducial cut efficiency 0.66 is unfolded for the ATLAS results.}
\label{fig:diphoton}
\end{figure}

The diphoton channel has long been recognized as a promising place to search for stoponium~\cite{Drees:1993yr}. 
At LO, the $\eta_{\tilde t}\rightarrow gg$ and $\eta_{\tilde t}\rightarrow \gamma\gamma$ partial decay widths do not depend on stop sector parameters other than the lightest stop mass, $m_{\widetilde t_1}$. They are induced solely by the strong and electromagnetic interactions, such that the ratio, 
\beq
\frac{\Gamma( \gamma \gamma )}{\Gamma( gg )} \simeq \frac{8 \alpha^2}{9 \alpha_s^2},
\label{eq:R}
\eeq
is  fixed by gauge couplings and independent of $m_{\widetilde t_1}$. As discussed in \Sec{sec:branching}, in our numerics we include the
NLO corrections to the ratio in Eq.~(\ref{eq:R})
from Ref.~\cite{Martin:2009dj,Younkin:2009zn}. These corrections, along with the running gauge couplings, induce a mild
dependence on the lightest stop mass, with the ratio~(\ref{eq:R})  varying from  $0.0025 - 0.0042$ as the lightest stop mass $m_{\tilde t_1}$ varies from 100 - 400 GeV. The NLO corrections actually reduce the ratio~(\ref{eq:R}) by 35\%--25\% from the LO result~\cite{Martin:2009dj}.

Since the $gg$ and $\gamma\gamma$ decay modes are always present and the size of the corresponding partial decay widths are rather model-independent, the diphoton branching ratio is maximized simply when the other possible decay channels ($WW$, $ZZ$, $hh$) are minimized. We will therefore define the {\it ideal $\gamma \gamma$ scenario} by the idealized limit in which the branching ratio ${\rm BR}(\eta_{\tilde t}\rightarrow \gamma\gamma)$ is given by the ratio in Eq.~(\ref{eq:R}) (including NLO corrections as discussed above). We note that the idealized branching ratios are $10^2$ -- $10^5$ times bigger than that of the SM Higgs boson with the same mass in the range of $m_{\eta_{\widetilde t}} = 200 - 600$ GeV.

In the full stop parameter space, the ideal diphoton scenario is approximately realized in the region where the Higgs-stop-stop coupling $\lambda_{h^0 \widetilde{t}_1 \widetilde{t}_1^*}$  is small, corresponding to mixing angles
\begin{equation}
\sin^2{\theta}_t \approx \frac{1}{2}\left( 1 \pm \sqrt{ 1 - \frac{4 m_t^2}{m_{\tilde t_2}^2 -m_{\tilde t_1}^2  }}  \, \right).
\end{equation}
In this region of parameter space the $\eta_{\tilde t} \rightarrow WW,ZZ,hh$ diagrams involving $s$-channel Higgs exchange are of a similar size or smaller than those mediated by the weak gauge couplings, thus explaining the dominance of the $\eta_{\tilde t} \rightarrow gg$ mode.
This can also be seen from the dependence of $\lambda_{h^0 \widetilde{t}_1 \widetilde{t}_1^*}$ and the stoponium branching ratios on the stop mixing angle in \Fig{fig:BR}, 
where we observe that the $gg$ branching ratio is nearly 100\% when the $\yhtt$ is small. For the stop mass $m_{\tilde t_1} = 100-300$ GeV ($m_{\tilde t_2} = 1$ TeV), the ideal $\gamma \gamma$ scenario is approximately realized for $\theta_t = 0.10-0.22$ or $\theta_t = 1.32-1.48$. 
The actual maximum diphoton branching ratio in these parameter regions is smaller than the ideal one of Eq.~(\ref{eq:R}) only by about $0.7-1.8\%$ fraction. In the next section, we will translate our limits to the full stop parameter space, accounting for the realistic branching ratios.

We note that the ideal $\gamma \gamma$ scenario lies within the blind spot region consistent with current indirect constraints, even for very light stop masses of order 100 GeV. In particular, the constraints from the Higgs signal strength data are relaxed in this region because the coupling $\lambda_{h^0 \widetilde{t}_1 \widetilde{t}_1^*}$ is small and the modification of the gluon-gluon fusion cross section from stop loops is minimized. This makes the $\eta_{\tilde t} \rightarrow \gamma\gamma$ channel particularly interesting to study as it can provide a complementary probe of this parameter region. 

In \Fig{fig:diphoton}, we show the current limits from ATLAS and CMS diphoton resonance searches and the future sensitivities for the ideal diphoton scenario. 
Stoponium masses $m_{\eta_{\tilde t}} \approx 2 m_{\tilde t_1}$
in the the vertical blue bands are excluded by the observed limits. 
We note that the ATLAS and CMS observed limits display complementary patterns of statistical fluctuations, excluding somewhat different stop mass ranges. The union of the two searches excludes stop masses up to about $m_{\widetilde t_1} \sim 136$ GeV. 
We emphasize that these exclusions are based on the assumption of including the first two $S$-wave stoponium states in the production rate, as described in \Sec{sec:prod}. This is a conservative assumption as it is likely that higher modes also contribute to the diphoton signal.  Following Ref.~\cite{Younkin:2009zn}, for comparison we have also displayed in \Fig{fig:diphoton} the cross section including only first $S$-wave mode or first 10 $S$-wave modes, allowing the reader to see the effect of this uncertainty. The reader can also see the difference between the two potential models: the thick blue line indicates the empirical quarkonia potential in Ref.~\cite{Hagiwara:1990sq} and the dot-dashed red line represents the Coulomb potential.

We also show in \Fig{fig:diphoton} the 14 TeV projections, based on the extrapolation of the current expected limits; see Appendx~\ref{app:extrap} for a description of the methodology behind our extrapolation. 
At the early stage of running with 30/fb, stops as heavy as about 180 GeV can be excluded in the ideal diphoton scenario, while a high luminosity run with 3/ab can exclude the stop up to about 370 GeV.

\subsection{$Z\gamma$} \label{sec:zgamma}

Another important channel is $Z\gamma$ since the branching ratio is also maximized when the coupling $\yhtt$ is small and the indirect constraints are absent. As for the diphoton channel, we can write the ratio of $Z \gamma$ to $gg$ partial widths at LO:
\beq
\frac{\Gamma(Z\gamma) }{ \Gamma(gg)} \, \simeq \, \frac{\alpha^2}{\alpha_s^2} \, \frac{1}{c_W^2 s_W^2} \left( c_t^2 - \frac{4}{3} s_W^2 \right)^2 \left(1- \frac{m_Z^2}{m_{\eta_{\tilde t} }}, \right).
\label{eq:zgammaratio} \eeq
As in the diphoton case, the BR($Z\gamma$) is maximized when the BR($gg$) dominates and the other decay modes ($ZZ,WW,hh$) are suppressed, 
corresponding to small $\yhtt$. 
However, unlike the diphoton case the mixing angle dependence in Eq.~(\ref{eq:zgammaratio}) implies that the BR$(Z\gamma)$ is largest when the lightest stop $\tilde t_1$ is  mostly left-handed. This asymmetry is also clearly visible in \Fig{fig:BR}. We therefore define the  
\emph{ideal $Z\gamma$ scenario} by evaluating BR($Z\gamma$) with stop parameters corresponding to $\yhtt=0$ and a mostly left-handed $\tilde t_1$. 
In practice, we choose different mixing angles satisfying this condition for different stop masses. 
The mixing angle and the BR($Z\gamma$) for the ideal $Z\gamma$ scenario gradually increase from $\theta_t = 0.169$ to $0.184$ and from 0.84\% to 1.37\%, respectively, as the stop mass varies in the range $m_{\widetilde t_1} = 100-400$ GeV. We note that these branching ratios are 100 -- 3000 times larger than those of the SM Higgs boson with the same mass in the range of $m_{\eta_{\widetilde t}} = 200 - 600$ GeV.

\begin{figure}[t] \centering 
\includegraphics[width=0.49\textwidth]{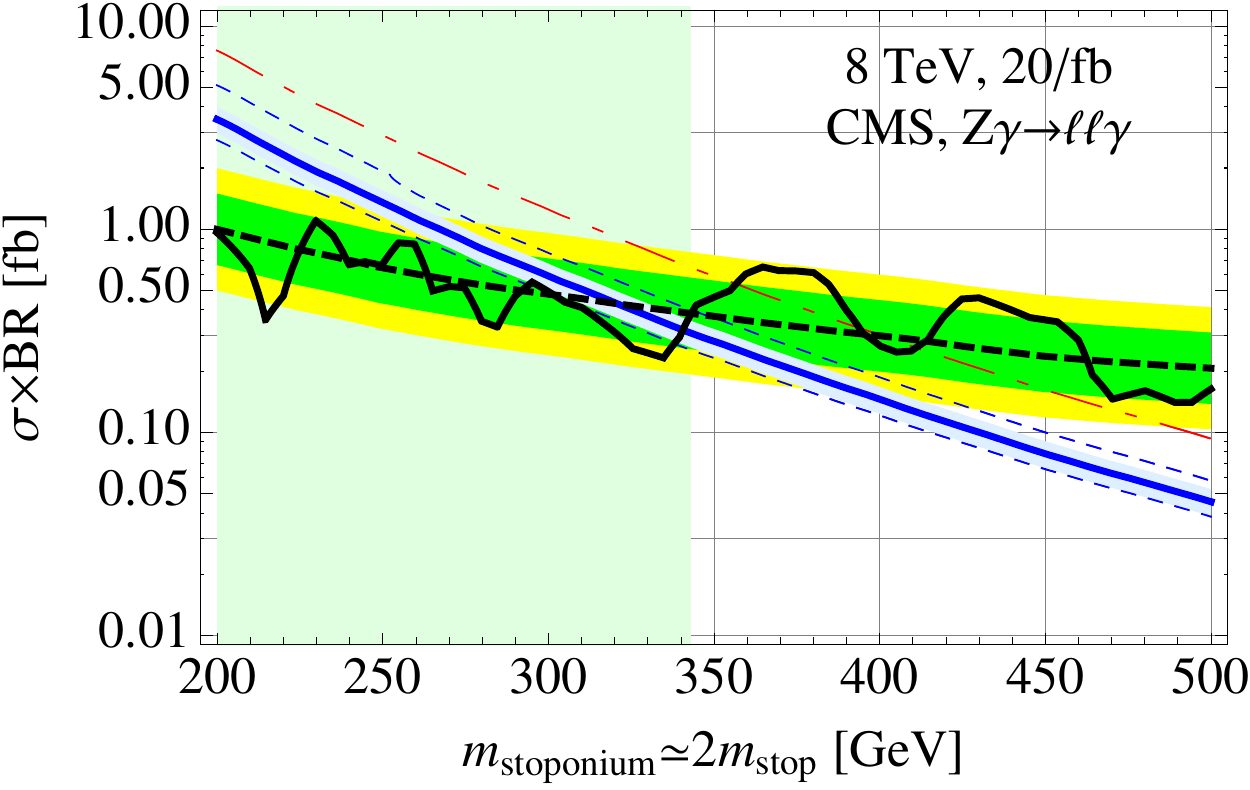}
\includegraphics[width=0.49\textwidth]{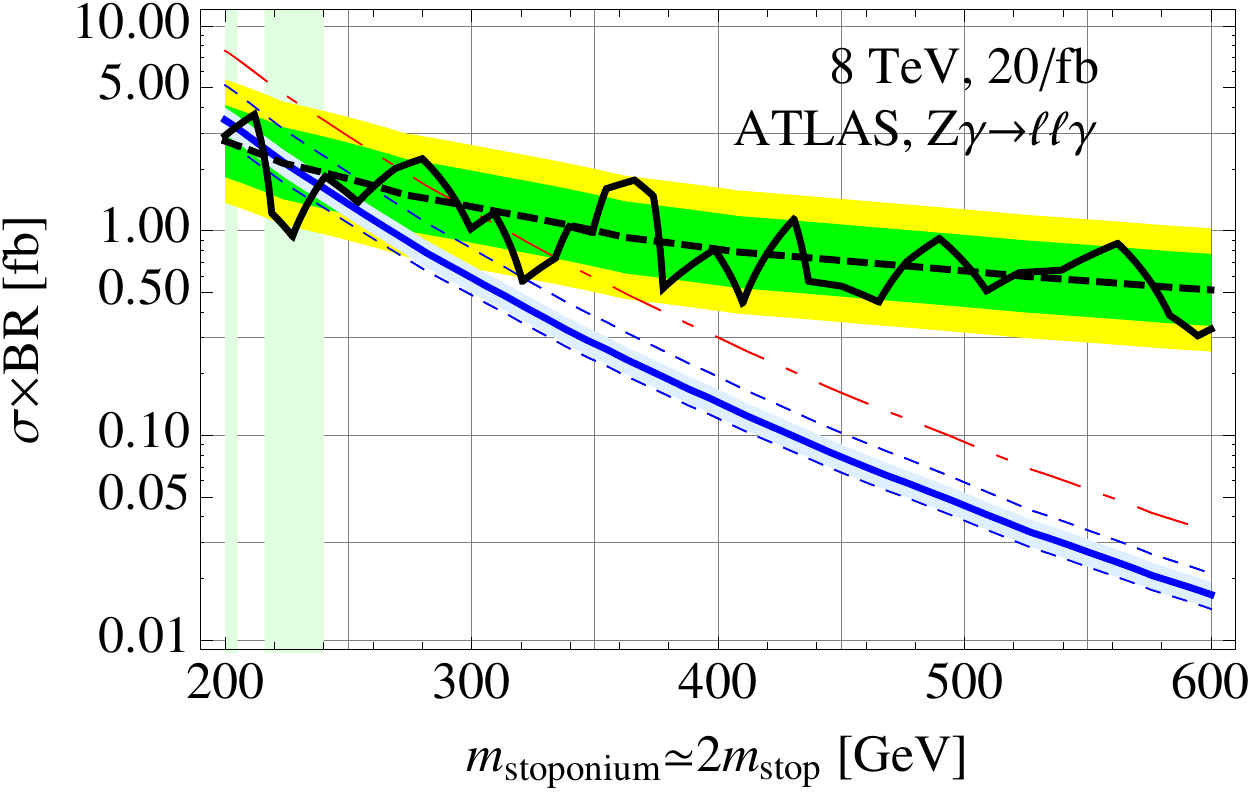}
\includegraphics[width=0.49\textwidth]{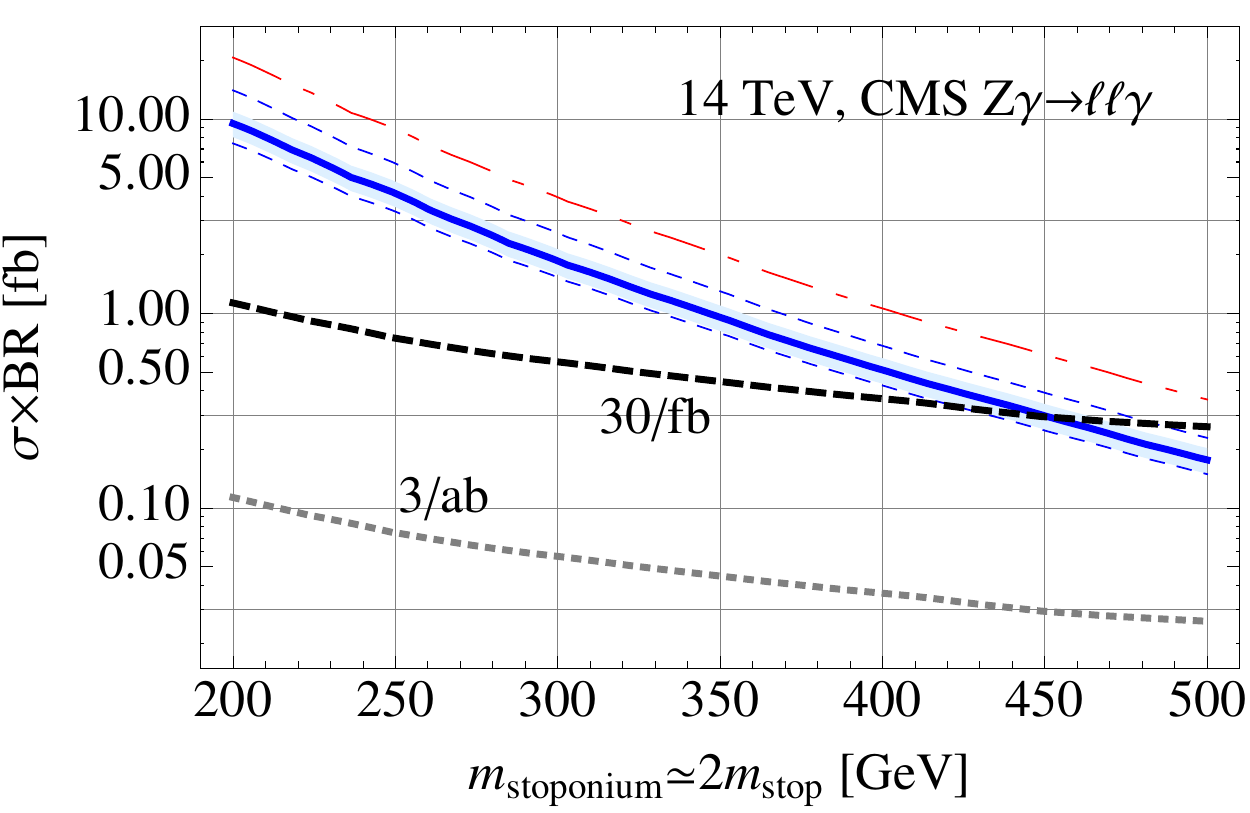}
\includegraphics[width=0.49\textwidth]{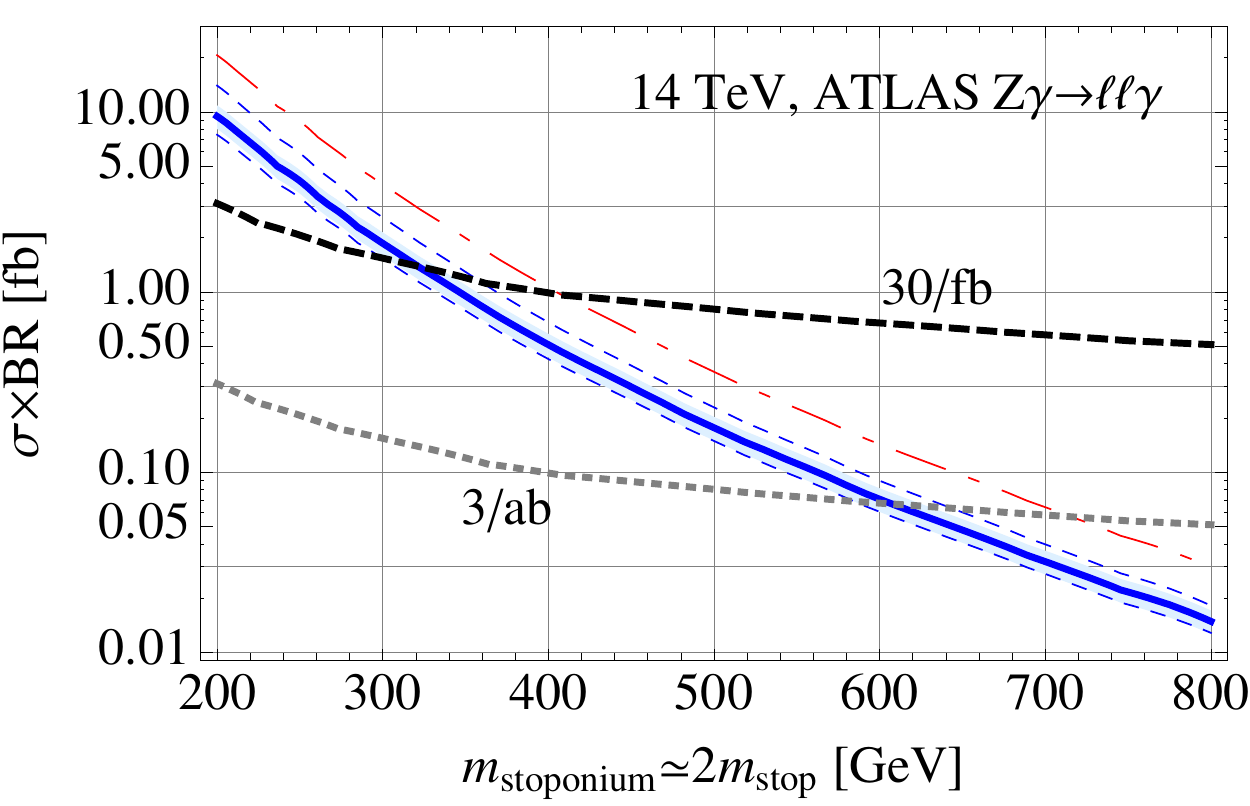}
\caption{\baselineskip=16pt 
Current bounds (upper) and future sensitivities (lower) on the ideal $Z \gamma$ scenario defined in \Sec{sec:zgamma} from ATLAS\,(right) and CMS\,(left) results. Theoretical predictions and excluded regions are as in \Fig{fig:diphoton}. A constant fiducial cut efficiency 0.6 is unfolded for ATLAS results.}
\label{fig:zgamma}
\end{figure}

We show the current limits on the ideal $Z\gamma$ scenario from the $Z \gamma \to \ell^+ \ell^- \gamma$ resonance searches in the upper panels of \Fig{fig:zgamma}. Notably, the CMS and ATLAS results differ somewhat in the excluded stop mass range; the CMS result is stronger and excludes stops up to about 170 GeV in the ideal $Z\gamma$ scenario, while instead the ATLAS result is sensitive only up to 120 GeV stops. In terms of the excluded cross-section the CMS result is about 2-3 times stronger for the stop mass range shown. Since the $Z\gamma$ result can potentially give the strongest constraint in the currently unconstrained parameter space with a mostly left-handed stop and small $\yhtt$, we encourage further dedicated analysis between the two experiments. 
We also show the 14 TeV prospects in the lower panels of \Fig{fig:zgamma}. 
The CMS-extrapolated result can potentially probe the stop up to 225 GeV with 30/fb and the ATLAS-extrapolated result can be sensitive to 300 GeV with 3/ab.
As we will see in \Fig{fig:summary} in the next section, the diphoton channel is most complementary to the indirect constraints, as it is sensitive to the entire small $\yhtt$ region including primarily right-handed stops. On the other hand, the $Z\gamma$ channel provides stronger limits than the diphoton channel for mostly left-handed stops.

It is important to mention that our $Z\gamma$ results are based on the LO calculation while the $\gamma \gamma$ result in the previous subsection is from the NLO calculation in Ref.~\cite{Martin:2009dj,Younkin:2009zn}. For the latter case, the NLO effects reduce the ideal $\gamma\gamma$ branching ratio by 35\%--25\% for the stop mass in the range 100-400 GeV. If the same NLO reduction is applied to the ratio of the $Z\gamma$ to $gg$ partial widths (hence, to the ideal BR($Z\gamma$)), the expected exclusion on the stop mass is approximately reduced by about 20 GeV. In addition to the NLO perturbative QCD effects, uncertainties from the number of $S$-wave stoponium states contributing to the signal and from potential models are again significant as shown by multiple theoretical predictions in \Fig{fig:zgamma}.

\subsection{$ZZ$ and $WW$}

The branching ratios BR($ZZ$) and BR($WW$) are intimately related with the coupling $\yhtt$ as discussed in \Sec{sec:branching}. The indirect probes of light stops discussed in \Sec{sec:interplay} place strong constraints on large values of this coupling and in turn determine the maximum phenomenologically allowed branching ratios BR($ZZ$) and BR($WW$), as is depicted in \Fig{fig:BR}. 

\begin{figure}[t] \centering 
\includegraphics[width=0.49\textwidth]{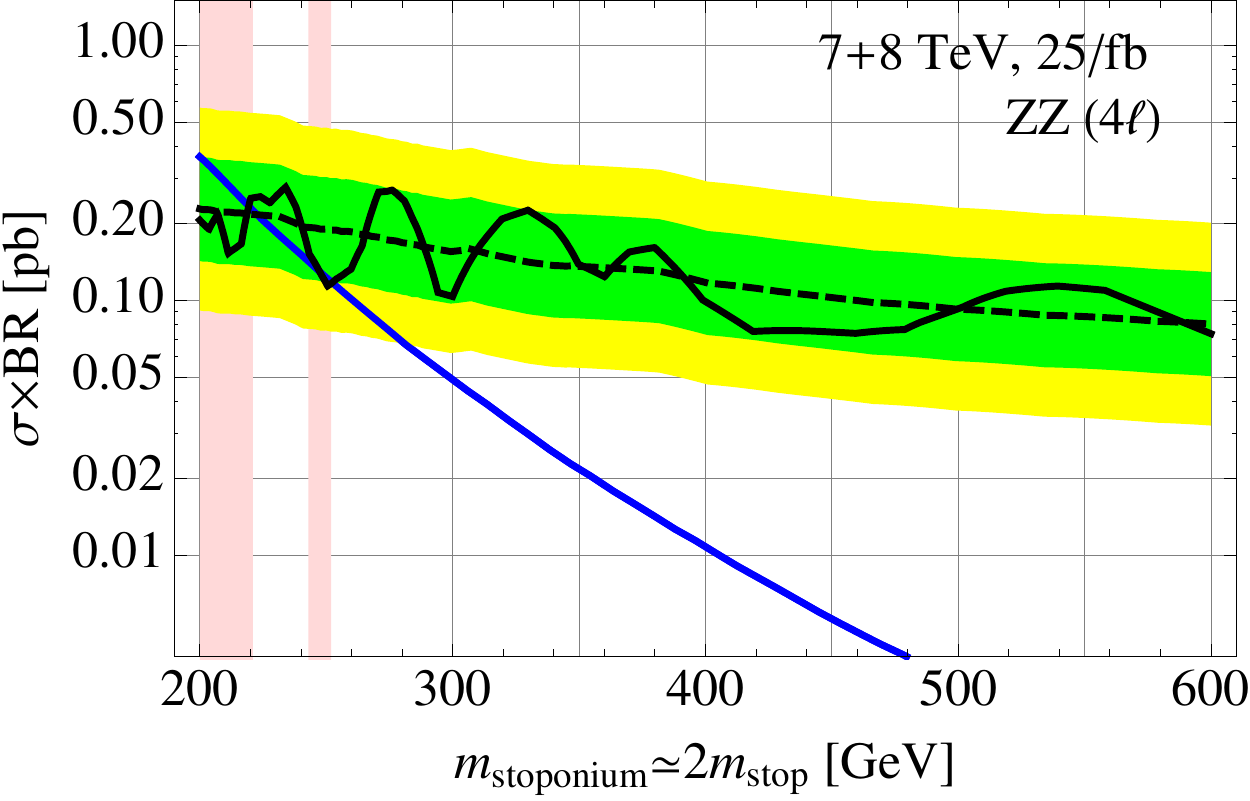}
\includegraphics[width=0.49\textwidth]{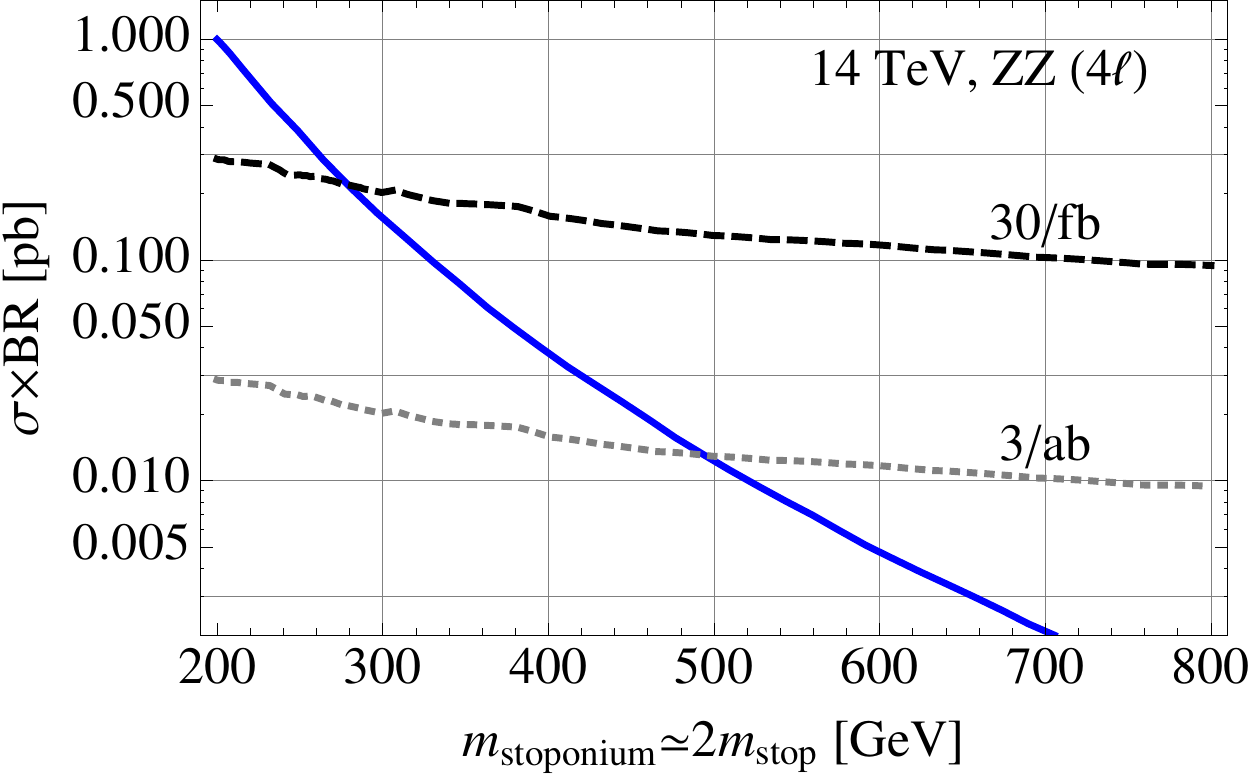}
\caption{\baselineskip=16pt 
Current(left) and future(right) bounds on the benchmark $ZZ$ scenario with BR($ZZ$)=0.06. CMS $ZZ \to 4\ell$ search result is used. }
\label{fig:ZZ}
\end{figure}

\begin{figure}[t] \centering 
\includegraphics[width=0.49\textwidth]{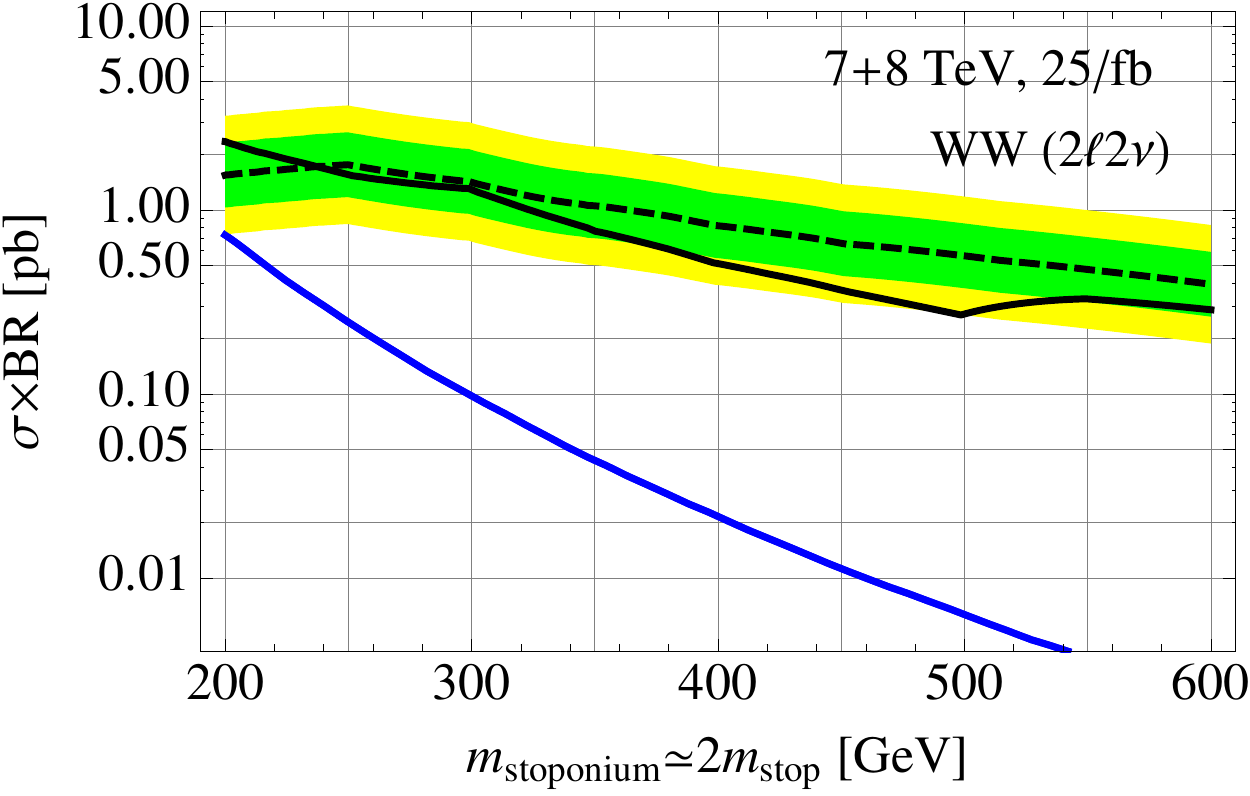}
\includegraphics[width=0.49\textwidth]{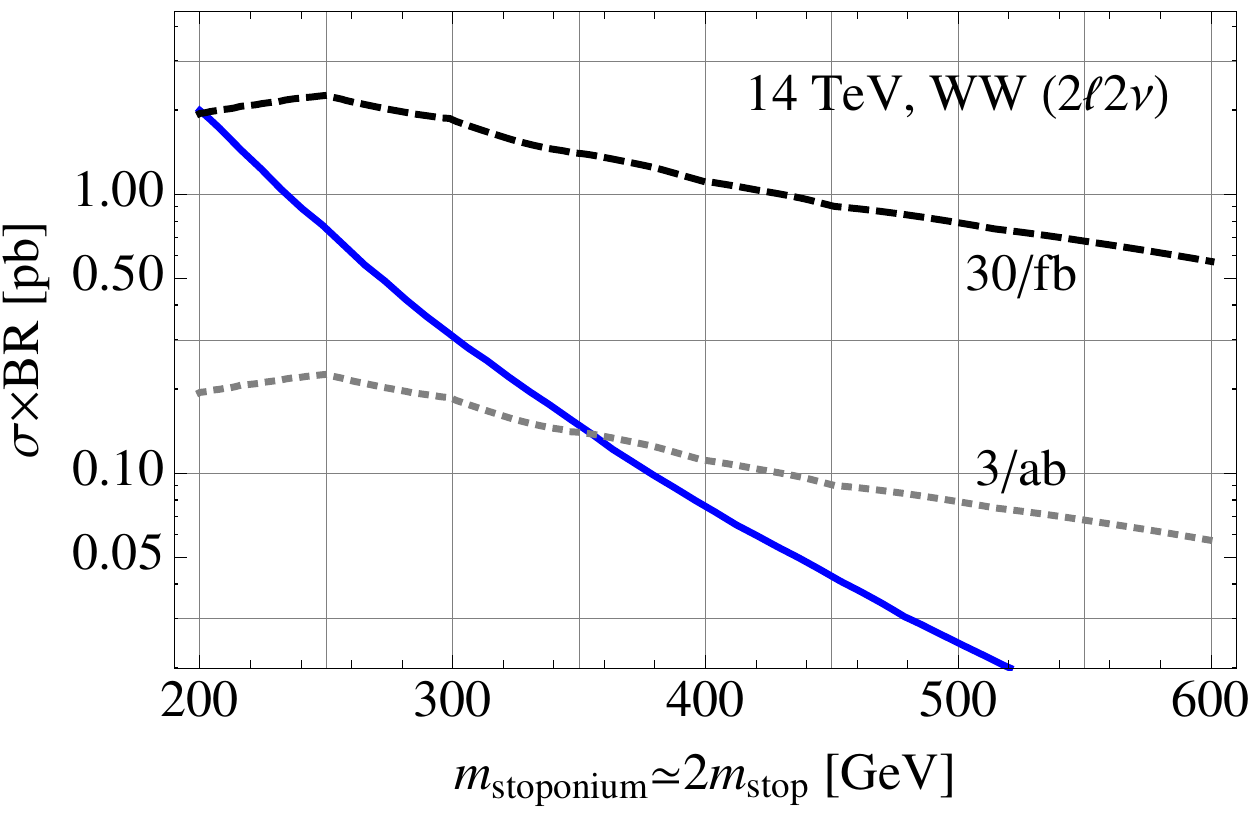}
\caption{\baselineskip=16pt 
Current(left) and future(right) bounds on the benchmark $WW$ scenario with BR($WW$)=0.12. CMS $WW \to 2\ell 2\nu$ search result is used. }
\label{fig:WW}
\end{figure}

Unlike the diphoton channel discussed above, there are several reasonable ways to define the benchmark $ZZ$, $WW$ scenarios. For instance, one could simply take the largest theoretically allowed branching ratio, or alternatively the largest branching ratio consistent with indirect constraints; in both cases one is led to vary the mixing angle as a function of the stop masses. One could also choose a fixed mixing angle for a range of the stop mass. For simplicity, we will instead choose a prescription with a fixed branching ratio that is broadly consistent with the current indirect constraints. Referring to \Fig{fig:BR} of $m_{\widetilde{t}_1}=160$ GeV case, we define the \emph{benchmark $ZZ$, $WW$ scenarios} by choosing BR($ZZ$)=6\% and BR($WW$)=12\%. In fact, for the range of $\theta_t$ allowed for the 160 GeV stop we obtain maximum BR($ZZ) \simeq$ 4 - 8\% and BR($WW) \simeq$ 11 - 16\% for the stop mass $m_{\widetilde t_1} = 100 - 300$ GeV (with $m_{\widetilde t_2} = 1$ TeV). Thus, the maximum branching ratios in this range of $\theta_t$ do not vary much with the stop mass. The current indirect constraints are somewhat weaker for heavier stops and a wider range of $\theta_t$ is viable, but we envisage that indirect constraints will also improve 
with more data, probing a broader range of mixing angles and heavier stop mass. We refer to the $m_{\tilde t_1} = 160$ GeV case as a representative benchmark scenario in view of the interplay between stoponium and indirect constraints. Again, we provide limits on the full stop parameter space accounting for true branching ratios in the next section.

Using a fixed branching ratio as a benchmark scenario for a wide range of ${\cal O}(100)$ GeV stop masses is also consistent with the softened Goldstone enhancement of the stop decay into longitudinal $Z$ and $W$ bosons, which could have sensitively increased the branching ratio with the stop mass. The decay into gauge bosons are maximized when the amplitude is dominated by the $s$-channel exchange of Higgs boson, corresponding to values of the $\yhtt$ coupling strong enough to dominate the decay but weak enough to be consistent with indirect constraints. But this contribution softens the Goldstone enhancement factor $(m_{\widetilde t_1}^2/ m_{W,Z}^2)$ of the amplitude by a suppression $1/(m_{\eta_{\widetilde t}}^2 - m_h^2) \approx 1/(4 m_{\widetilde t_1}^2)$ from the propagator, leaving the dominant amplitude relatively insensitive to the stop mass for the benchmark parameter space.

In \Fig{fig:ZZ} and \Fig{fig:WW} we show the current limits and prospects of the benchmark $ZZ$ and $WW$ scenarios from the $ZZ \to 4 \ell$ and the $WW \to 2\ell 2\nu$ channels, respectively. The constraint from the $ZZ$ channel is stronger, just as it is for the SM Higgs. The $WW \to \ell \nu 2j$ channel gives similar bounds as the $2 \ell 2 \nu$ channel but is not shown for simplicity. Currently, the $ZZ$ channel constrains stops up to about 110 GeV and a small range of 121-126 GeV for the benchmark scenario. In the future, the $ZZ$ channel is expected to probe 140 (250) GeV stops at $\sqrt{s} = 14$ TeV  with 30/fb (3/ab) for the $ZZ$ benchmark scenario. 
While there are no limits yet, the LHC at 14 TeV with 3/ab is expected to probe 180 GeV stops in the benchmark $WW$ scenario. We note that in the full stop parameter space, with realistic branching ratios, there are in fact new limits from the $WW$ channel for very light stops, where the $hh$ mode is absent.

\subsection{$hh$}

Similarly to the $ZZ$ and $WW$ branching ratios, the maximum branching ratio Br$(hh)$ is limited by indirect constraints on the coupling $\yhtt$. We define the  \emph{benchmark $hh$ scenario} by a fixed BR($hh$)=0.25. The actual maximal BR($hh$) for the range of $\theta_t$ currently allowed for the 160 GeV stop gradually increases from 16\% to 41\% for the 130 GeV to 300 GeV stop.

\begin{figure}[t] \centering 
\includegraphics[width=0.49\textwidth]{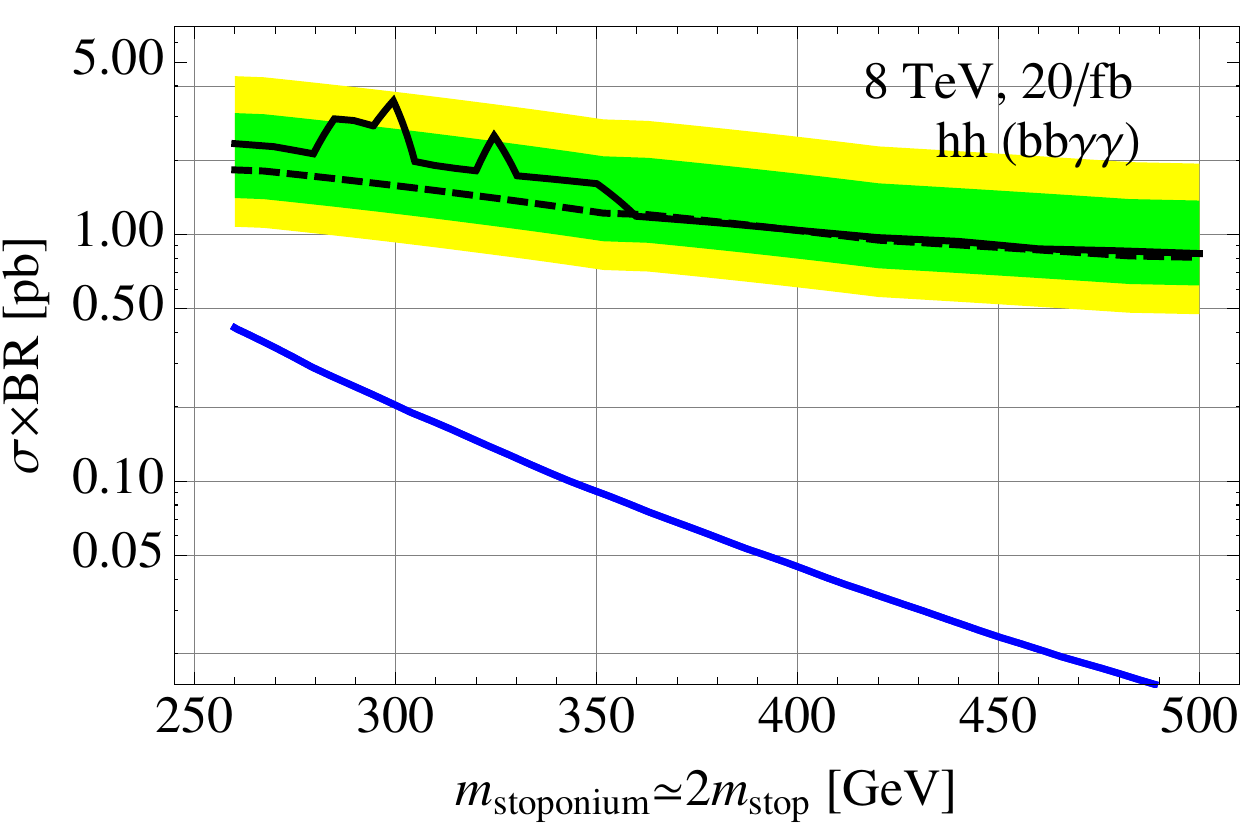}
\includegraphics[width=0.49\textwidth]{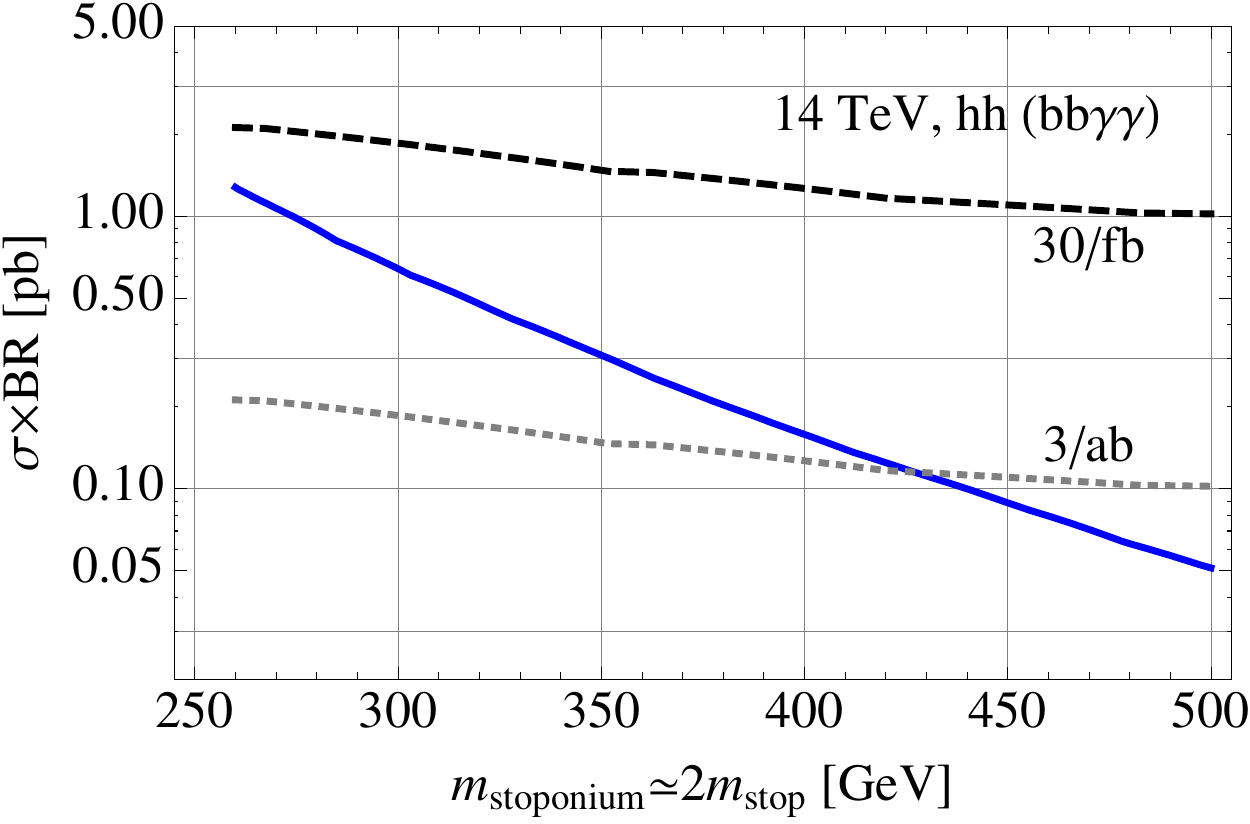}
\caption{\baselineskip=16pt 
Current(left) and future(right) bounds on the benchmark $hh$ scenario with BR($hh$)=0.25. ATLAS $hh \to b\bar{b} \gamma \gamma$ search result is used.}
\label{fig:hh}
\end{figure}

In \Fig{fig:hh}, we display the current sensitivity and 14 TeV projections on the benchmark $hh$ scenario from the $hh \to b\bar{b} \gamma \gamma$ resonance search. No new limits can be obtained with this channel from the Run 1 data. However, in the long term this channel is expected to probe stops as heavy as 215 GeV in the benchmark $hh$ scenario at $\sqrt{s} =14$ TeV with 3/ab. 

\begin{figure}[t] \centering 
\includegraphics[width=0.92\textwidth]{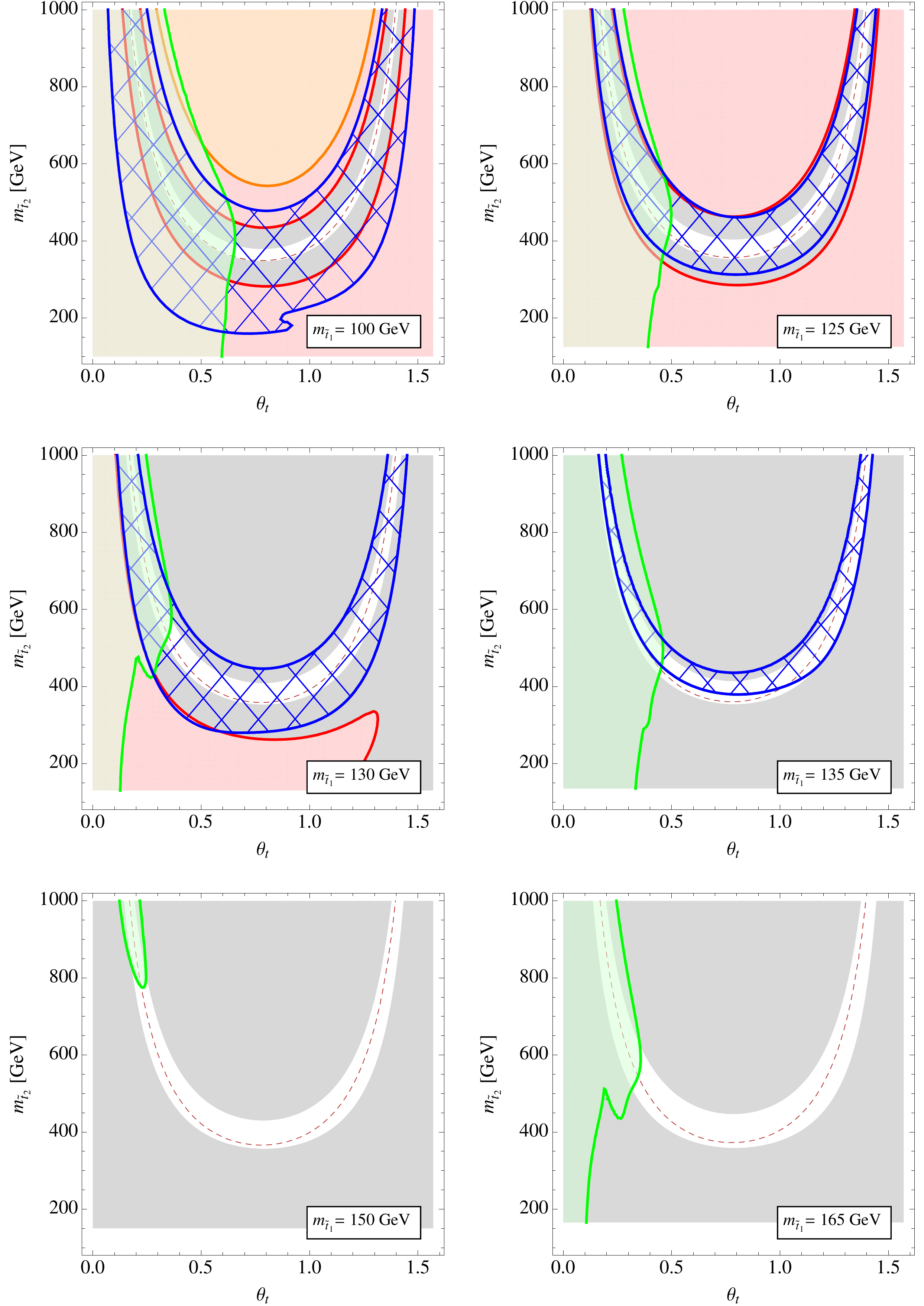}
\vspace{-0.1in}\caption{\baselineskip=16pt 
\emph{Excluded regions in the physical stop parameter space.} Here we show the limits from diboson resonance searches in the $\theta_t$-$m_{\tilde t_2}$ plane for fixed $m_{\tilde t_1} = (100,125,130,135,150, 160)$ GeV, including $\gamma \gamma$ (blue hatched), $Z \gamma \to \ell^+ \ell^- \gamma$ (green), $ZZ \to 4 \ell$ (red), $WW \to 2\ell 2\nu$ (orange). We also display the union of indirect constraints (gray) shown in \Fig{fig:indirect}. 
The strongest limit from ATLAS or CMS in each channel are shown. Finally, we indicate the stop parameters for which $\yhtt=0$ (red dashed contour).}
\label{fig:summary}
\end{figure}

\section{Summary: interplay of stoponium and indirect constraints}
\label{sec:interplay}

Our main results are presented in \Fig{fig:summary}, which displays the constraints from diboson resonance searches on stoponium in the physical stop parameter space. We show six slices of the parameter space, fixing the lightest stop mass $m_{\tilde t_1} = (100,125,130,135,150, 160)$ GeV and showing the limits in the $\theta_t$-$m_{\tilde t_2}$ plane. The individual constraints from the $\gamma\gamma$, $Z\gamma$, $ZZ$, and $WW$ channel are represented (no bounds from $hh$ channel can be obtained from the current data).  For comparison we also display the region excluded by indirect constraints coming from Higgs signal strength measurements, electroweak precision data, and vacuum  stability (see Appendix \ref{sec:indirect}). In contrast to the previous section in which we defined ideal or benchmark branching ratios, here we account for the realistic branching ratios, computed at each point in the physical stop parameter space as described in \Sec{sec:branching}. 

Assuming the stoponium forms and annihilate decays, we observe that light stops with masses $m_{\tilde t_1} \lesssim 130$ GeV are excluded by the combination of diboson resonance searches and the indirect constraints. Stoponium searches alone probe stops lighter than about 125 GeV. Remarkably, a strong complementarity is seen between the $\gamma\gamma$ resonance searches and the indirect constraints (particularly the Higgs coupling constraints), with the $\gamma\gamma$ channel probing the remaining open stop blind spot region corresponding to small or vanishing coupling $\yhtt$. 
Moreover, we also see that primarily left-handed stops with masses $m_{\tilde t_1} \lesssim 170$ GeV are constrained by $Z\gamma$ resonance searches.  

Again, we emphasize that the limits we present here rely on the $\Lambda_{\overline {\rm MS}}^{(4)} = 300$ MeV parameterization of the stoponium wavefunction at the origin from Ref.~\cite{Hagiwara:1990sq} and the inclusion of the first two $S$-wave states~\cite{Younkin:2009zn}. While it is likely that this estimate is conservative, particularly in light of the possible additional signal coming from the higher excited states, it is worthwhile to examine how the limits change if different assumptions are made. For example, if we instead include only the contribution of the 1S state, the expected limit on the stoponium mass will be weaker by about 20 GeV, as can be seen by examining Figs.~\ref{fig:diphoton} and
\ref{fig:zgamma}. Alternatively, if one assumes the pure Coulomb potential, the expected limit on the stoponium mass will be stronger by about 70 GeV. In any case, it is clear that stoponium is now probing the hypothesis of light stops beyond the LEP limits and will explore uncharted territory with the next run of the LHC.

\section{Example scenarios of interest for stoponium}
\label{sec:scenarios}

We now discuss two motivated examples scenarios in which stoponium provides a complementary probe to direct stop searches and indirect tests. 
The condition~(\ref{eq:cond}) will generically hold when the stop does not possess an unsupressed 2-body decay. This can happen due to kinematics or small couplings, and to demonstrate this we will focus two examples - 1) a ``canonical" case of R-parity conservation with a bino LSP and a stop NLSP, and 2) a R-partiy violating case with a stop LSP which decays to a pair of jets through the $UDD$ operator. 

\subsection{R-Parity conserving bino LSP and stop NLSP} 

\begin{figure}[t] \centering 
\includegraphics[width=0.47\textwidth]{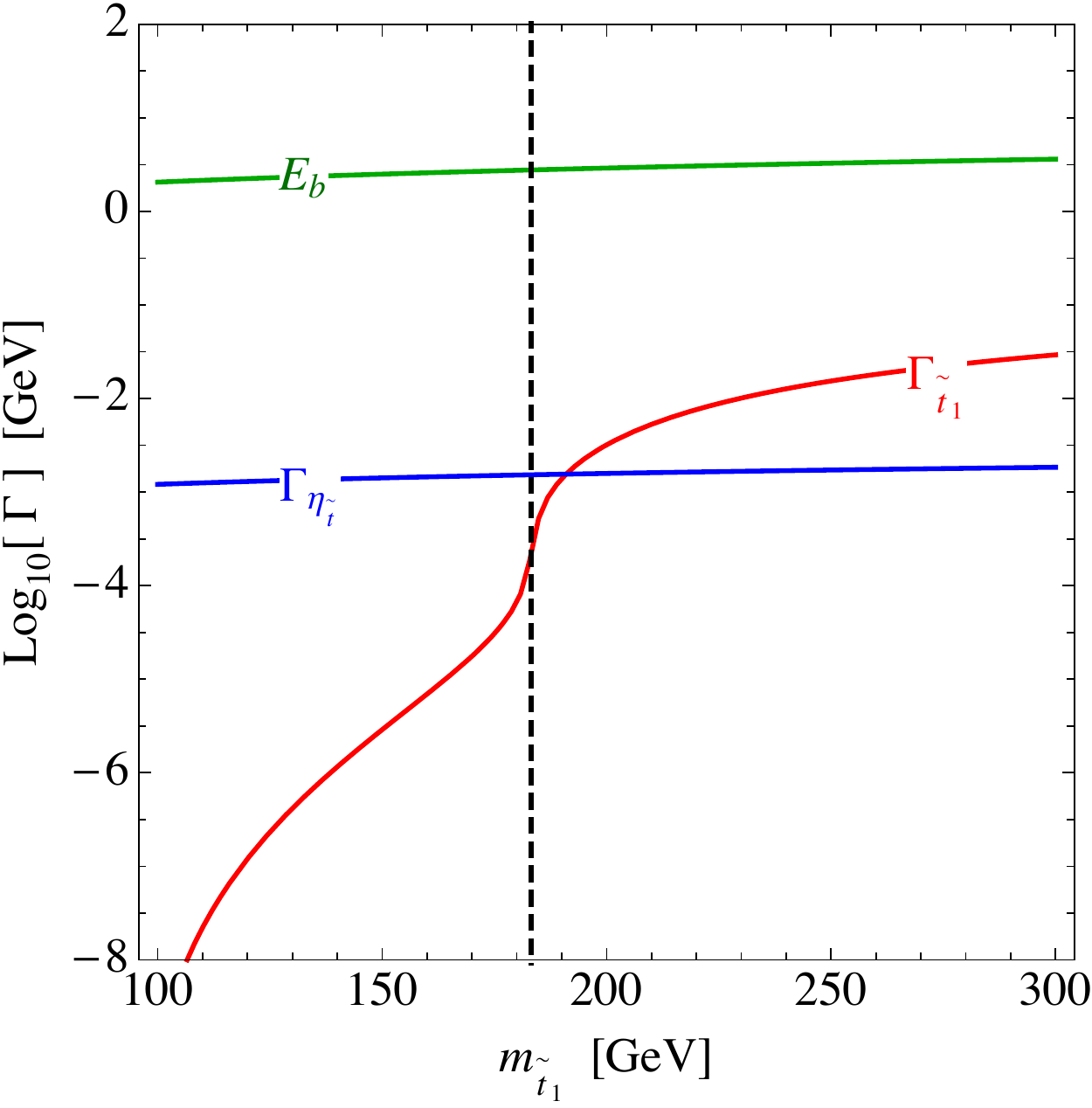}
\includegraphics[width=0.5\textwidth]{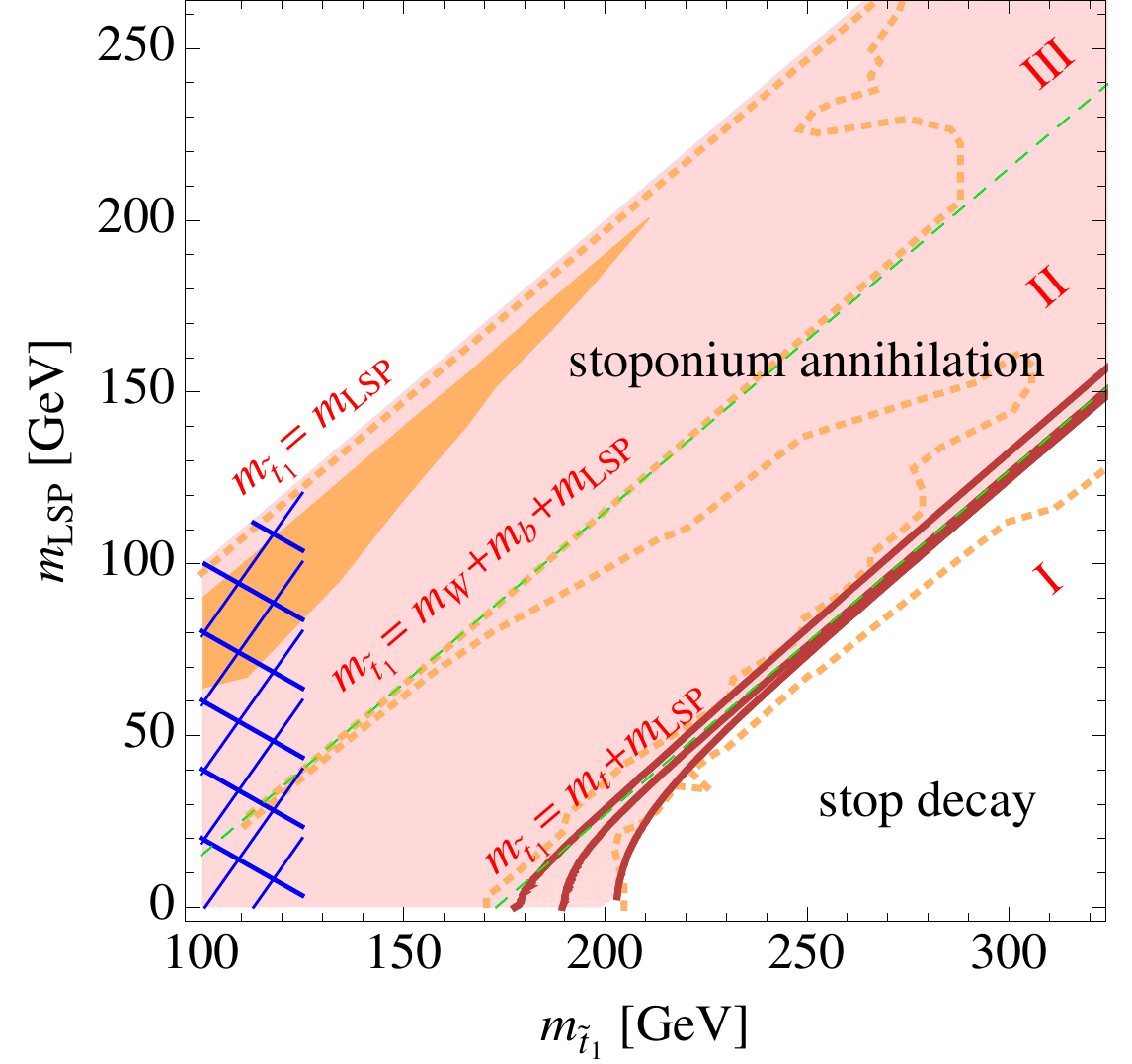}
\caption{\baselineskip=16pt 
{\it R-parity conserving bino LSP and stop NLSP}: {\bf (Left):} Comparison of the stop decay width\,(red), stoponium annihilation rate\,(blue) and stoponium binding energy\,(green). Here we have assumed a pure bino LSP with $M_1=10$ GeV, $m_{\tilde t_2} = 1$ TeV, and $\theta_t = 0.18$. The vertical line separates the two- and three-body decays of the stop. 
{\bf (Right):} In the red shaded region above each red line, the stoponium annihilation is quicker than the individual stop decay, and searches for stoponium resonances are possible. Three benchmark scenarios are presented as red lines: the top two have almost ideal $\gamma \gamma$ branching ratios (defined in \Sec{sec:diphoton}) with $\theta_t \approx 1.36$\,(top), 0.18\,(middle), and the bottom one with $\theta_t = 0.12$\,(bottom). For comparison, we also display a union of ATLAS limits on the stops  (orange dotted), which assume 100\% branching ratios to the final states under consideration~\cite{ATLAS:stopsummary} as well as a branching ratio-independent exclusion limit taken from Ref.~\cite{Grober:2014aha} (orange shaded). The blue-hatched region is excluded from stoponium~\Fig{fig:summary}.}
\label{fig:stform}
\end{figure}

A canonical benchmark scenario with light stops consists of a stop NLSP and a neutralino LSP, under the assumption that R-parity is conserved. In this case the neutralino is stable, leading to the signature of missing energy at colliders. As is well known, in this scenario the stop width is very sensitive to the mass spectrum, and several kinematic regions can be distinguished: 
\begin{enumerate}[I.]
\item $m_{\tilde t_1} > m_t + m_{\chi^0}$:  The stop decays via the 2-body process $\tilde t_1 \rightarrow t \chi^0$.
\item $m_W + m_b + m_\chi^0 < m_{\tilde t_1} < m_t + m_\chi^0$: The stop decays via the 3-body process $\tilde t_1  \rightarrow W b \chi^0$.
\item $m_b + m_\chi^0 <  m_{\tilde t_1} < m_W + m_b + m_\chi^0$: The stop decays via the 4-body process $\tilde t_1 \rightarrow b f  f' \chi^0$, or possibly through the flavor-violating 2-body decay $\tilde t_1 \rightarrow c \chi^0$.
\end{enumerate}
To illustrate these features, in the left panel of \Fig{fig:stform} we have plotted the decay width of the stop as a function of its mass for the case of a 10 GeV bino LSP (see the Appendix~\ref{app:decay} for the 3-body partial decay width of the stop). For comparison, in the same plot we show the stoponium binding energy and annihilation decay width. As the stop mass approaches the boundary between 2- and 3-body decays, we observe that $\Gamma_{\eta_{\widetilde t}}$ becomes larger than the natural stop width. Therefore, in region I above, the stop generally decays too quickly for stoponium resonance signatures to be visible, although it may be possible near the edge of region I where phase space suppression is relevant. However, in regions II and III the dominant stop decay channels are 3- and 4-body (or suppressed 2-body), respectively, allowing both sufficient time for the formation of the stoponium and the dominance of the bound state annihilation decays. 

Therefore, regions II and III are prime targets for resonance searches for stoponium, which is clearly illustrated by the ``phase'' diagram in the stop -- neutralino mass plane shown in the right panel of \Fig{fig:stform}. We represent the current exclusion limit from stoponium resonances by the blue hatched region; this region is excluded for the given lightest stop mass for all other stop parameters (see \Fig{fig:summary}). For comparison, we have overlaid the current ATLAS limits from direct stop searches~\cite{ATLAS:stopsummary}, represented by the orange dotted line. While these direct limits naively appear to cover most of light stop region, it is important to emphasize that they rely on the assumption of a 100$\%$ branching ratio to the final state under consideration. If this assumption is relaxed, the limits are weakened considerably. For example, in region III if the stop has comparable branching ratios in the 4-body $b f f' \chi^0$ and 2-body $c \chi^0$ channels, a significant portion of parameter space opens up~\cite{Grober:2014aha}, and only the orange-shaded region can be excluded regardless of the assumption on the branching ratio. 
In contrast, constraints from stoponium resonances are already starting to unambiguously probe this region.

It is also worthwhile to note that the ``stealth'' stop region -- the phase space-suppressed part of region I near $m_{\widetilde t_1} \sim m_t$, $m_{\chi^0} \gtrsim 0$ -- is also potentially amenable to stoponium searches (within the red-shaded region in the bottom part of \Fig{fig:stform}) 
This region is challenging to probe with direct stop searches, and several dedicated observables such as spin correlations~\cite{Han:2012fw,Aad:2014mfk}, top pair production rates~\cite{Czakon:2014fka,Aad:2014kva} and special kinematic variables~\cite{Kilic:2012kw} have been studied. However, they primarily rely on precision measurements and calculations, which are often subject to subtle uncertainties. 
On the other hand, the stoponium resonance searches are cleaner and less ambiguous in most cases. 
Thus, the stoponium can provide an important complementary probe of the stealth region. 
In this region, stops with a mostly left-handed mixture (small $\theta_t$) can be probed via stoponium over a wider range of stop masses (see \Fig{fig:stform} right) because the individual stop decay width is smaller with a smaller $\theta_t$. 
In any case, stoponium does not yet have sensitivity to stealth stops (the blue-hatched region extends to about 130 GeV), but can potentially be sensitive to a significant portion of this challenging region in the early part of Run 2 with $\sim {\cal O}(10)$/fb of data (see \Fig{fig:diphoton}).

Let us finally comment on the case in which the LSP is a very weakly interacting particle such as the gravitino.
In this scenario, the stop NLSP is typically long lived due to its weak coupling to the gravitino, and therefore stoponium annihilation decays can be relevant. 
For instance,
taking a low supersymmetry breaking scale $\sqrt{F}=10$ TeV (corresponding to an essentially massless gravitino), a 200 GeV stop has a decay width $\Gamma(\widetilde{t_1} \to t \widetilde{G}) \sim 10^{-9}$ GeV~\cite{Sarid:1999zx}, which is much smaller than the typical stoponium annihilation width.
In practice, however, this scenario is already strongly constrained by direct searches covering both prompt and displaced decays~\cite{Liu:2015bma}, and only a small window around the stealth stop parameter region for prompt decays is still open. The stoponium searches will be able to fill this open region. 

\subsection{RPV stop LSP}

\begin{figure}[t] \centering 
\includegraphics[width=0.49\textwidth]{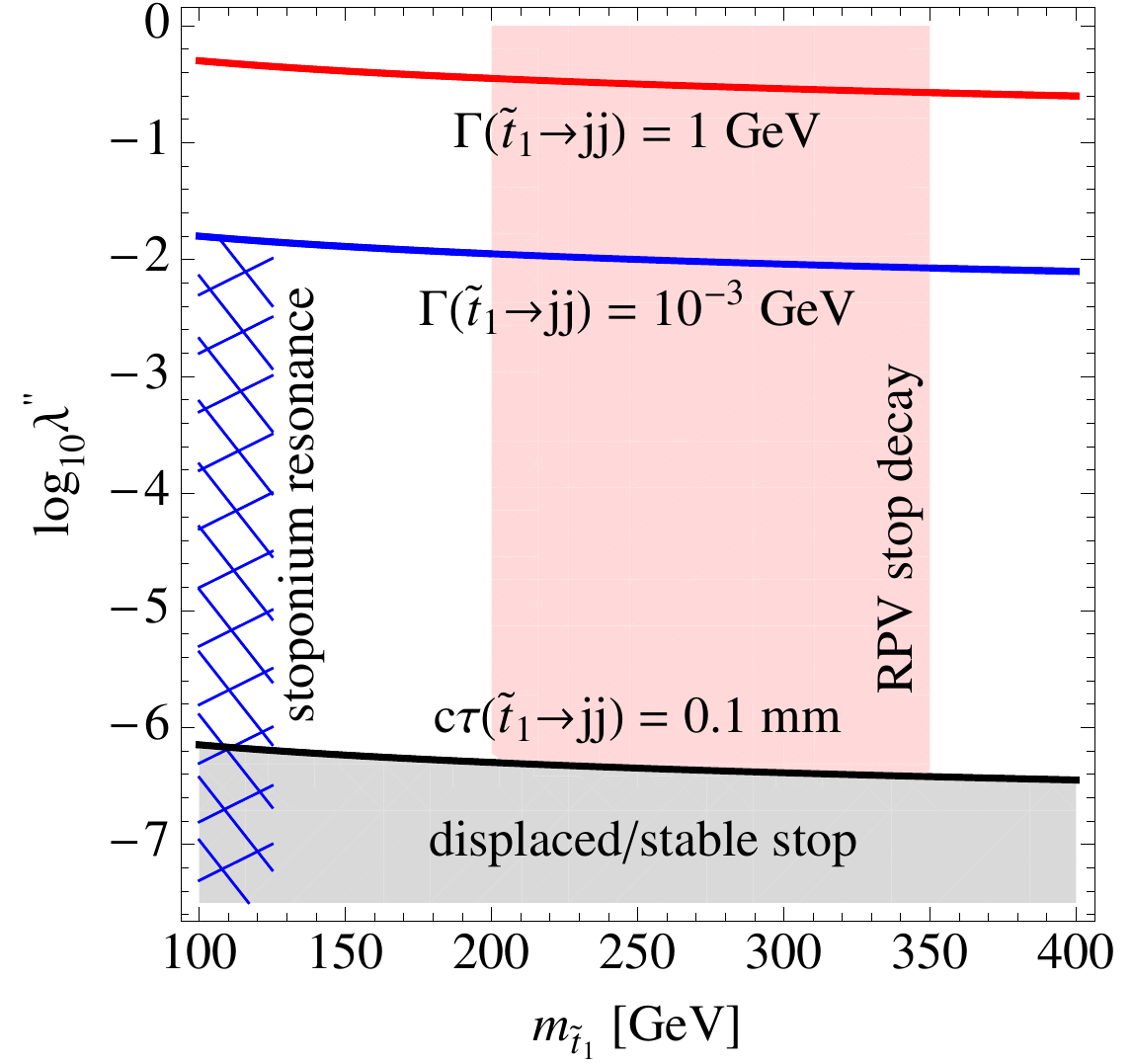}
\caption{\baselineskip=16pt 
{\it R-parity violating stop LSP}: We display contours for two values of the stop width, $\Gamma_{\tilde t_1} = 1$ GeV (red) and $\Gamma_{\tilde t_1} = 10^{-3}$ GeV (blue) for typical values of stoponium binding energies and annihilation decay widths. We also display a contour of the stop lifetime $c \tau_{\tilde t_1} = 0.1$ mm. Stops with lifetimes in this range and longer are subject to stringent LHC searches for long-lived charged particles~\cite{ATLAS:2014fka,Liu:2015bma}. Stops with prompt RPV decays in the red-shaded region are also constrained from RPV searches~\cite{ATLAS:2012ds}. Stoponium can therefore be relevant for a broad range of RPV couplings, $\lambda^{''} \lesssim 10^{-2}$, and the current exclusion from stoponium resonances is shown as blue-hatched.  }
\label{fig:decayrpv}
\end{figure}

Another benchmark scenario of interest is a stop LSP that decays to a pair of jets due to a small $UDD$ R-parity violating coupling, which we customarily denote as $\lambda''$. 
Such a RPV stop is challenging to probe with direct searches at LHC. 
Despite its large production cross section, the signature of four jets (paired dijet resonances) is difficult to disentangle from the QCD background, and stops as light as ~$\sim 100$ GeV are still allowed~\cite{Evans:2012bf,Bai:2013xla,ATLAS:2012ds} although some range of the stop masses $200-350$ GeV has recently been 
excluded~\cite{Khachatryan:2014lpa}.

Stoponium resonance searches provide another means of probing this interesting scenario. Since the decay of the stop is through a 2-body process, the RPV coupling must be somewhat suppressed in order for the  condition~(\ref{eq:cond}) to apply. In fact, this may be motivated in explicit models of R-parity violation which aim to protect the lifetime of the proton through other symmetries. For example, in the scenario of MFV SUSY~\cite{Csaki:2011ge,Nikolidakis:2007fc}, the dominant RPV coupling is 
$\lambda^{''}_{tsb} \sim \order(10^{-4})$, which is small enough for~(\ref{eq:cond}) to hold. Note also that if the RPV coupling is extremely small, $\lambda'' \lesssim \order(10^{-7})$ then the stop will have a macroscopic lifetime. In this case, there are additional strong constraints from searches for displaced dijets~\cite{CMS:2014wda,Liu:2015bma} or long-lived colored/charged particles~\cite{ATLAS:2014fka,Liu:2015bma}. In~\Fig{fig:decayrpv} we display the RPV stop LSP phase diagram. We conclude that stoponium resonance searches can provide new complementary probe of this scenario for a broad range of couplings, $10^{-7} < \lambda^{''} < 10^{-2}$.

\section{Outlook}
\label{sec:outlook}

A natural supersymmetric solution to the hierarchy problem predicts light stops, which can lead to a variety of novel phenomena. Besides the suite of direct searches at the LHC (which depend on the SUSY spectrum) and indirect tests such as Higgs couplings, precision electroweak data and vacuum stability requirements, 
it is possible that light stops can manifest through a stoponium annihilation decay leading to a resonance signature in the $\gamma \gamma,Z\gamma,ZZ,WW$ and $hh$ channels. 
This will happen if the stop does not have an unsuppressed two body decay and naturally occurs in several motivated scenarios. 

In this work we have derived new limits on light stops from ATLAS and CMS diboson resonance searches using Run 1 data. Our limits are derived using the empirical quarkonia potential model of~\cite{Hagiwara:1990sq} and assuming the first two $S$-wave states contributing to the resonance signal, which we have argued is conservative. Assuming the stoponium can form and annihilate decay, light stops below about 130 GeV are excluded by the combination of constraints on stoponium resonances and other indirect constraints, such as Higgs couplings measurements. Notably, we have demonstrated that the $\gamma\gamma$ channel is especially complementary to these indirect probes, as the resonance signature in this channel is maximized in the blind spot region where the Higgs-stop-stop trilinear coupling is small and these constraints are weakest. Furthermore, $Z\gamma$ resonance searches provide even stronger limits for primarily left handed stops, limiting stop masses below about 170 GeV. In the long term, the LHC experiments will be able to probe stops in the 300-400 GeV range with a high-luminosity run with 3/ab at 14 TeV.

We have also discussed the implications of these searches for two specific SUSY scenarios with lights stops. In the case of a bino LSP and stop NLSP, stoponium annihilation decays can be relevant since the stop width is naturally suppressed over a wide range of parameters. Stoponium resonance searches provide a relatively clean and unambiguous probe, and can give a handle on the challenging compressed and stealth stop regions of parameter space. Instead for the case of a R-parity violating stop LSP decaying to two jets, stoponium already provides unique sensitivity to stops lighter than about 130 GeV and RPV couplings in the range $10^{-7}<\lambda'' < 10^{-2}$, which is difficult to probe directly due to the large QCD backgrounds. 

Looking forward, it is clearly of interest on the theoretical side to bring under better control the uncertainties coming from our imprecise understanding of the stoponium bound state system. A modern study of empirical quarkonia potential models along the lines of Ref.~\cite{Hagiwara:1990sq} is warranted, and perhaps lattice studies could provide further insight into the non-perturbative matrix elements entering into stoponium production and decays. A more detailed investigation into the decay patterns of the excited states would also be helpful in order to understand their contribution to the resonance signals and perhaps uncover additional signatures of stoponium. 
On the experimental side, stoponium clearly provides a well motivated target for resonance searches, which should be a high priority during the next run of the LHC.

\subsubsection*{\bf Acknowledgements}

We thank Roberto Franceschini, Seyong Kim, Adam Leibovich, Florian Staub, Brock Tweedie and Yeo Woong Yoon for helpful discussions. We are also grateful to Guillelmo Gomez Ceballos, Louis Helary, Elisabeth Petit, Marco Pieri, Pierre Savard, and Livia Soffi for helpful discussions and correspondence regarding the ATLAS and CMS analyses. BB is supported in part by a CERN COFUND  fellowship. 
SJ is supported in part by the National Research Foundation of Korea under grant 2013R1A1A2058449. SJ thanks the Mainz Institute for Theoretical Physics and CERN theory group where some of this work was done.

\appendix
\numberwithin{equation}{section}

\section{Stop and sbottom sector conventions} \label{app:param}

In terms of the gauge eigenstates $(\tilde t_L, ~\tilde t_R)$, the stop mass matrix is given at tree level by
\begin{equation}
\left( 
\begin{array}{cc}
m_{Q_3}^2 + m_t^2 + D^t_L
& m_t X_t\\
m_t X_t  &  m_{U_3}^2 + m_t^2 + D^t_R
\end{array}
\right), 
\label{eq:stop-mass}
\end{equation}
where $m^2_{Q_3}$, $m^2_{U_3}$ are the left- and right-handed squark soft mass parameters, $X_t \equiv A_t -\mu/\tan\beta$, with $A_t$ the soft trilinear coupling, $\mu$ the supersymmetric Higgs mass parameter, and $\tan \beta$ the ratio of up and down type Higgs vacuum expectation values, and
$D^t_L =   m_Z^2 \cos{2 \beta} \left( \tfrac{1}{2}  -\tfrac{2}{3} s_W^2 \right)$, 
$D^t_R  =    m_Z^2 \cos{2 \beta} \left(   \tfrac{2}{3} s_W^2 \right)$.
For simplicity, we will assume all parameters are real.  The physical stop mass eigenstates are related to the gauge eigenstates through the orthogonal transformation:
\begin{equation}
\label{stopX}
\left(
\begin{array}{c}
\tilde t_L \\
\tilde t_R
\end{array}
 \right)
 = 
\left(
\begin{array}{cc}
 \cos \theta_{t} & -\sin\theta_{t} \\
\sin \theta_{t} &  \cos \theta_{t}
\end{array}
 \right)
\left(
\begin{array}{c}
\tilde t_1 \\
\tilde t_2
\end{array}
 \right),
\end{equation}
where the mixing angle $\theta_t$ satisfies
\begin{equation}
\tan 2 \theta_{t} = \frac{ 2 m_t X_t}{m_{Q_3}^2- m_{U_3}^2 + D^t_L - D^t_R},
\end{equation}
which we take in the range $[0, \pi/2]$.

The sbottom sector is described in a similar fashion. The sbottom mass matrix can be obtained from Eq.~\ref{eq:stop-mass} with the replacements 
$m_t \rightarrow m_b$, 
$m_{U_3}\rightarrow m_{D_3}$,  
$X_t \rightarrow X_b = A_b - \mu \tan \beta$, 
$D_L^t \rightarrow D_L^b =    m_Z^2 \cos{2 \beta} \left( -\tfrac{1}{2}  +\tfrac{1}{3} s_W^2 \right)$, and 
$D_R^t \rightarrow D_R^b =  m_Z^2 \cos{2 \beta} \left(  - \tfrac{1}{3} s_W^2 \right)$.

\section{Indirect constraints on light stops}
\label{sec:indirect}
\begin{figure}[t] \centering 
\includegraphics[width=0.4\textwidth]{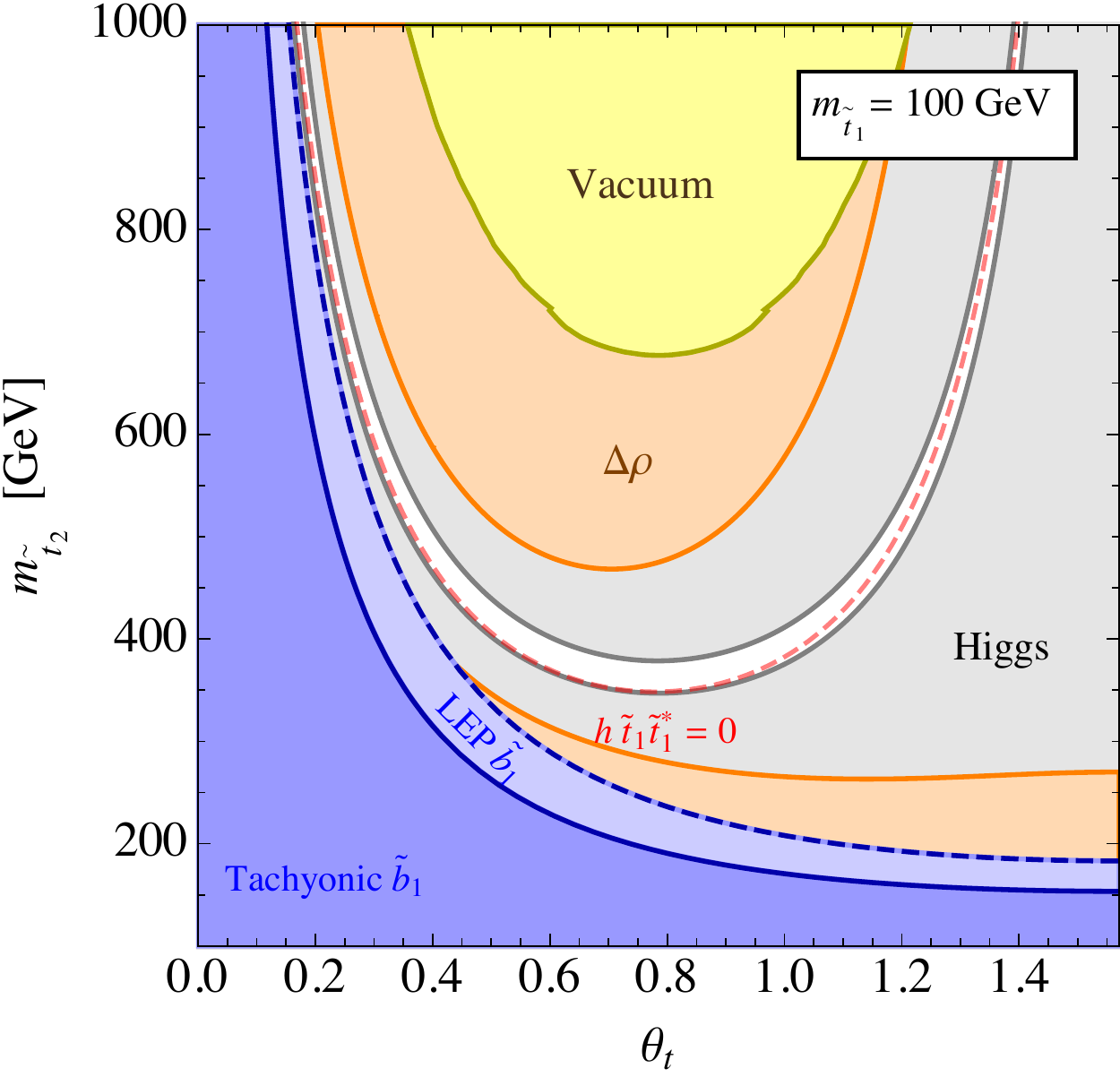} ~
\includegraphics[width=0.4\textwidth]{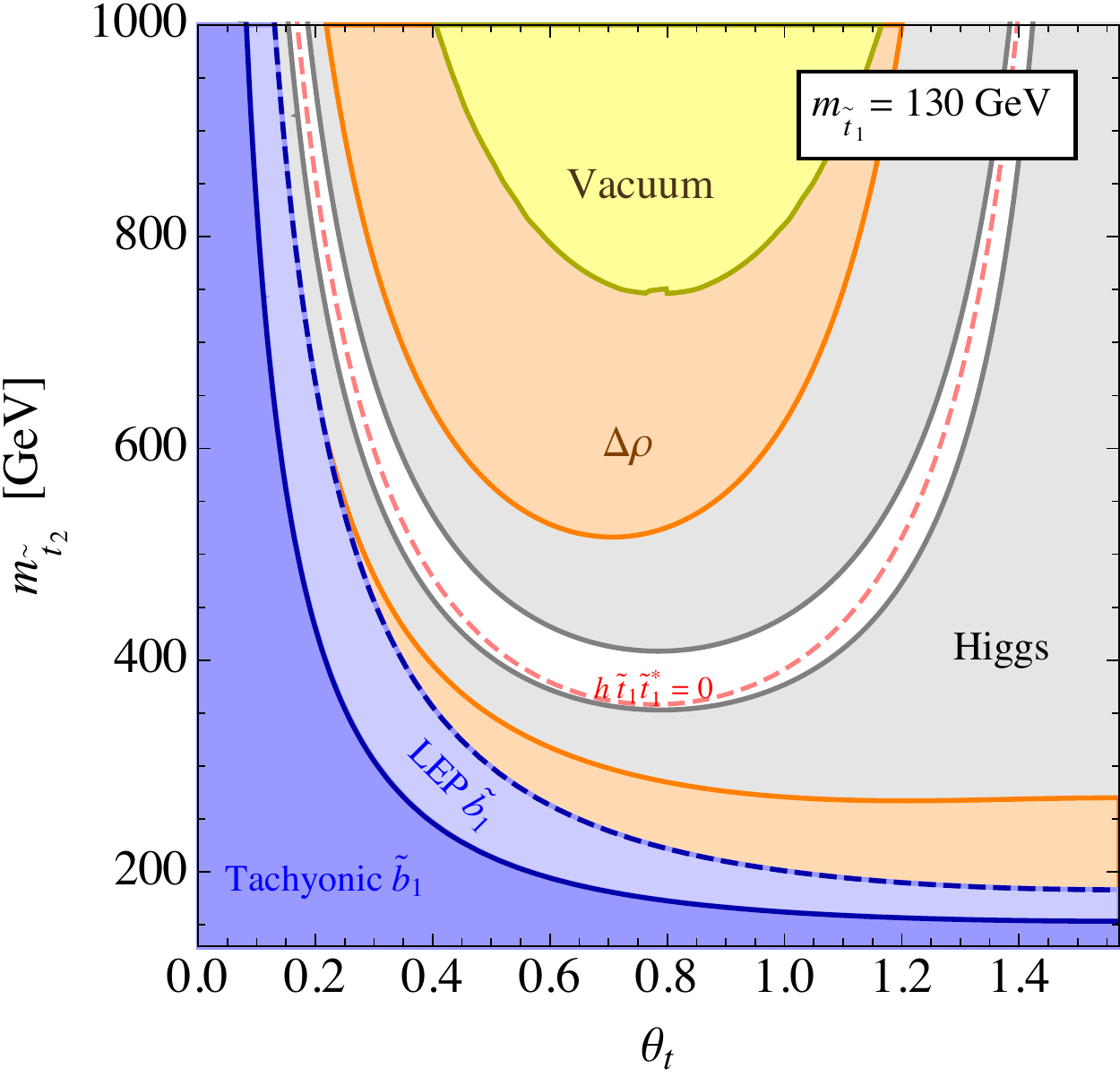}\\
\includegraphics[width=0.4\textwidth]{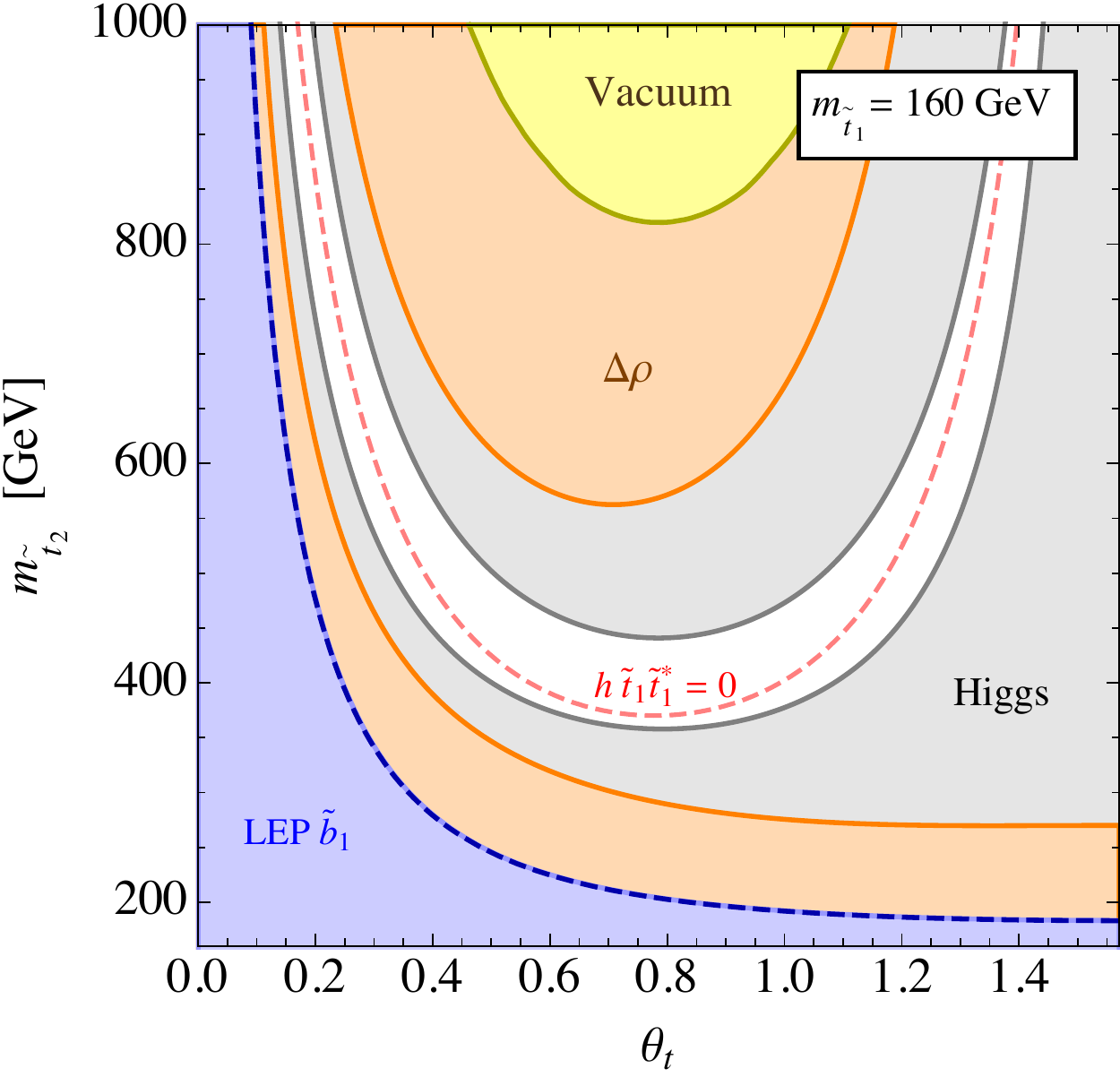} ~
\includegraphics[width=0.4\textwidth]{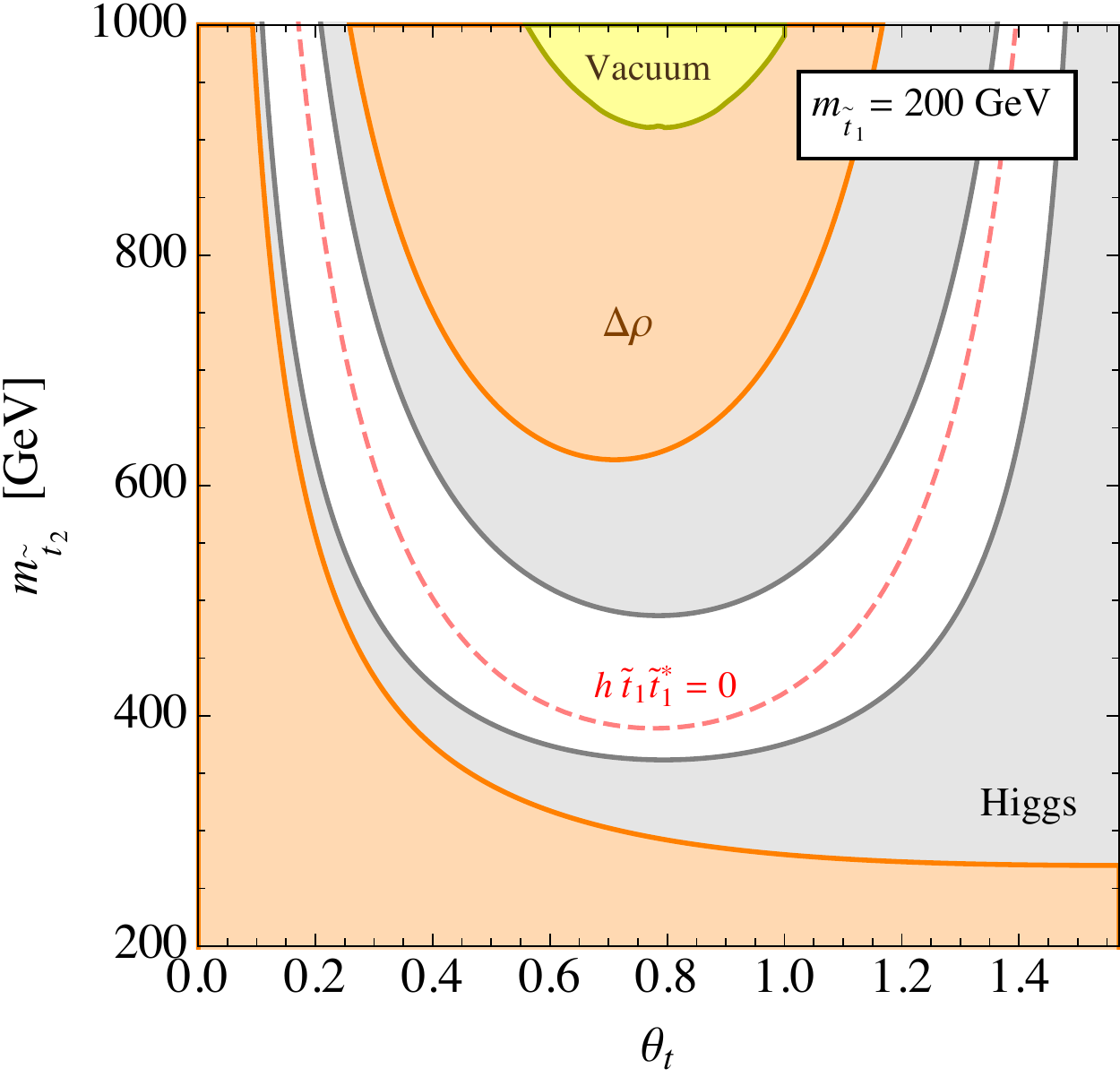}\\
\includegraphics[width=0.4\textwidth]{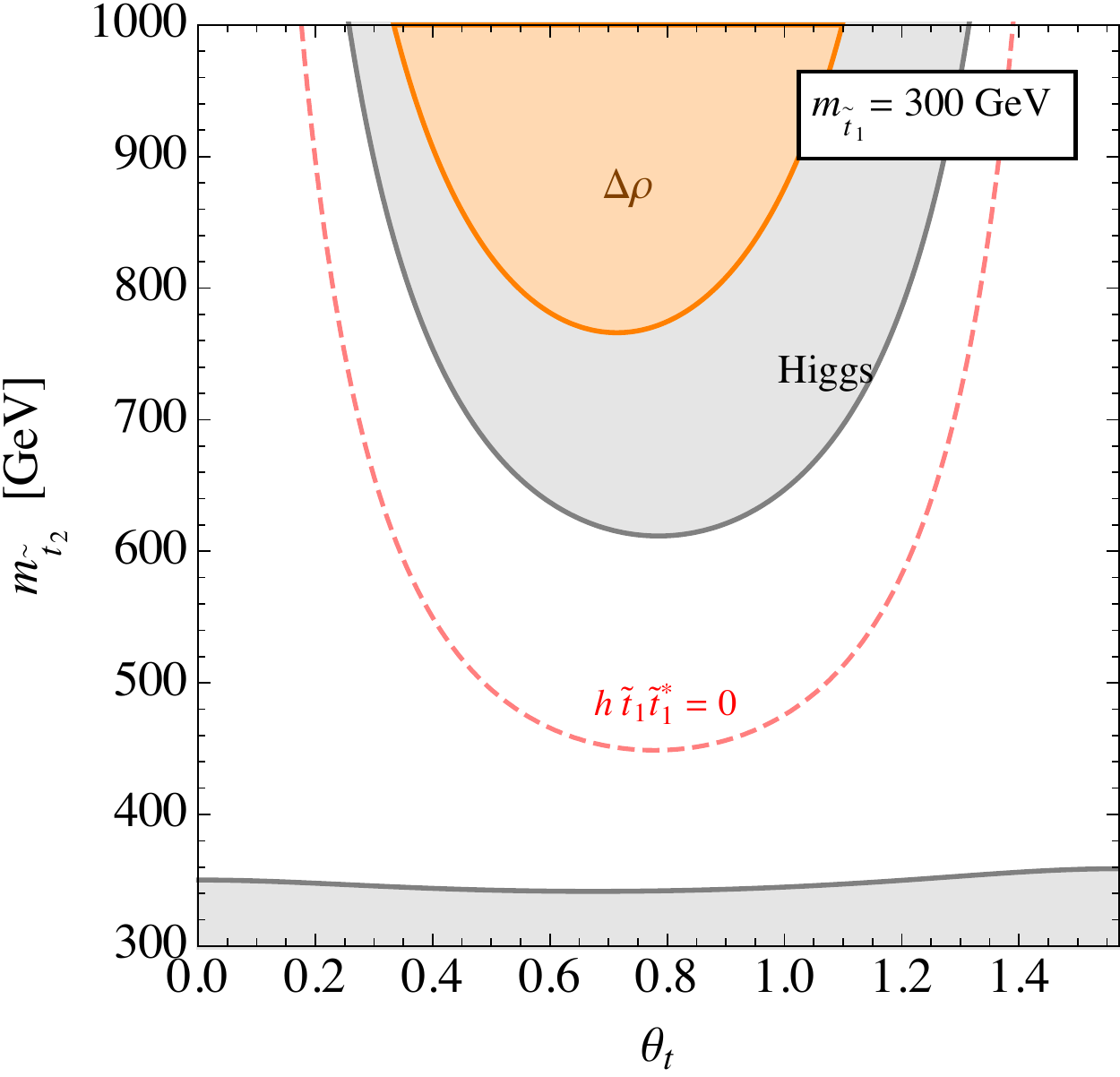}
\caption{\baselineskip=16pt 
 {\it Indirect constraints on light stops}:  Here we display for light stop masses $m_{\tilde t_1}$ = (100,130,160,200,300) GeV the constraints in the $\theta_t$ - $m_{\tilde t_2}$ plane coming from Higgs signal strength measurements (gray), precision electroweak data (orange), vacuum stability (yellow), tachyonic sbottoms (blue) and sbottoms below the LEP kinematic reach (light blue). The white regions are currently unconstrained. The red dashed line indicates where the Higgs-stop-stop coupling $\yhtt$ vanishes. Here we have fixed $\tan \beta = 10$ and the sbottom mixing to zero ($X_b = 0$).} 
\label{fig:indirect}
\end{figure}

In this appendix we summarize the existing indirect constraints on light stops coming from Higgs signal strength measurements, precision electroweak data, and vacuum stability. 
In the MSSM, there will be additional constraints on the stop parameters coming from the requirement of obtaining a 125 GeV Higgs mass, but we will remain open to new contributions to the Higgs quartic coming from physics beyond the MSSM. Furthermore, if Higgsinos or gluinos are light there can be further constraints from flavor physics, which we will not include here. 
Several summary plots of these constraints are shown in \Fig{fig:indirect}.

\subsection{Higgs signal strength data}
Light stops modify the $hgg$ and $h\gamma\gamma$ couplings at one loop and are therefore subject to constraints by ATLAS and CMS measurements of the Higgs signal strength parameters~\cite{higgs-stops}. To derive the exclusions, we use the method of Ref.~\cite{Belanger:2013xza}, 
with the recent updates in Ref.~\cite{Bernon:2014vta}, which include the latest Run 1 results as of summer 2014. 
In this approach, the raw signal strength data are combined to derive a $\Delta \chi^2$ function for the 
$\gamma\gamma$, $VV$, and $b/\tau$ channels as a function of two combined production mode signal strengths: 1) gluon-gluon fusion and $t th$ (ggF+ttH), and 2) vector boson fusion and associated production (VBF+VH). This yields eight composite signal strengths, $\hat \mu^{(\rm ggH+ttH,VBF+VH)}_{(\gamma\gamma,VV,b\bar b,\tau \bar \tau)}$. In terms of the ratios $r_i \equiv \Gamma(h \rightarrow i)/ \Gamma(h \rightarrow i)_{\rm SM}$, the signal strength predictions are given by
\begin{eqnarray}
& \displaystyle{ \hat \mu^{\rm ggH +ttH}_{\gamma\gamma} = \frac{r_{gg} \, r_{\gamma\gamma}}{r_\Gamma} } , 
~~~~~~~~~ \displaystyle{ \hat \mu^{\rm VBF + VH}_{\gamma\gamma} = \frac{r_{\gamma\gamma}}{r_\Gamma}}, & \nonumber \\
& \displaystyle{ \hat \mu^{\rm ggH +ttH}_{VV} = \hat \mu^{\rm ggH +ttH}_{b \bar b}
= \hat \mu^{\rm ggH +ttH}_{\tau \bar \tau}  = \frac{r_{gg} }{r_\Gamma}   } , &  \nonumber \\
& \displaystyle{ \hat \mu^{\rm VBF+VH}_{VV} = \hat \mu^{\rm VBF+VH}_{b \bar b}
= \hat \mu^{\rm VBF+VH}_{\tau \bar \tau}  = \frac{1}{r_\Gamma}   } , & 
\end{eqnarray}
where we have defined $r_\Gamma = \Gamma_h/\Gamma_h^{\rm SM} \simeq 1+(r_{gg} -1){\rm BR}(h\rightarrow gg)_{\rm SM}$ with BR$(h\rightarrow gg)_{\rm SM} \simeq 0.085$.

One loop sbottom and stop exchange lead to the following expressions for $r_{gg},r_{\gamma\gamma}$:
\begin{eqnarray}
\label{eq:rg}
r_{gg} &=& 
\Bigg\vert  \,
1 \, +
 \, \frac{\sqrt{2}\, v \, C(r)}{A_{gg}^{\rm SM}} \,
  \bigg[  
    \lambda_{h^0 \tilde t_1 \tilde t^*_1}  \frac{A_0(m_{\tilde t_1})}{ m_{\tilde t_1}^2  } + 
    \lambda_{h^0 \tilde t_2 \tilde t^*_2}  \frac{A_0(m_{\tilde t_2})}{ m_{\tilde t_2}^2  }   \\
    &&~~~~~~~~~~~~~~~~~~~~~~~~~~~~~~~~~~~~~~~~~~\,  + 
  \lambda_{h^0 \tilde b_1 \tilde b^*_1}  \frac{A_0(m_{\tilde b_1})}{ m_{\tilde b_1}^2  } + 
   \lambda_{h^0 \tilde b_2 \tilde b^*_2}  \frac{A_0(m_{\tilde b_2})}{ m_{\tilde b_2}^2  }  
   \bigg]
\Bigg\vert^2,   \nonumber \\
\label{eq:rga}
r_{\gamma\gamma}  &=&   
\Bigg\vert  \,
1 \, - \,
 \frac{  v  \, d(r) }{\sqrt{2} A_{\gamma \gamma }^{\rm SM}}  \,
  \bigg[  Q_{\tilde t}^2 \left(  
   \lambda_{h^0 \tilde t_1 \tilde t^*_1}  \frac{A_0(m_{\tilde t_1})}{ m_{\tilde t_1}^2  } + 
    \lambda_{h^0 \tilde t_2 \tilde t^*_2}  \frac{A_0(m_{\tilde t_2})}{ m_{\tilde t_2}^2  } \right) \\
      & & ~~~~~~~~~~~~~~~~~~~~~~~~~~~~~~~~~ \,+  
  Q_{\tilde b}^2 \left(  
   \lambda_{h^0 \tilde b_1 \tilde b^*_1}  \frac{A_0(m_{\tilde b_1})}{ m_{\tilde b_1}^2  } + 
    \lambda_{h^0 \tilde b_2 \tilde b^*_2}  \frac{A_0(m_{\tilde b_2})}{ m_{\tilde b_2}^2  } \right)    
   \bigg]
\Bigg\vert^2,  \nonumber 
\end{eqnarray}
where $C(r)=1/2$, $d(r)=3$, $Q_{\tilde t}=2/3$, $Q_{\tilde b}=-1/3$, and 
$A_{gg}^{\rm SM} \approx 1.3$, $A_{\gamma\gamma}^{\rm SM} \approx 6.6$. The Higgs-squark trilinear couplings $\lambda_i$ can be found in, \eg, Ref.~\cite{Martin:2002iu}.
The scalar loop function $A_0(m)$ is defined in Ref.~\cite{Djouadi:2005gj} and
approaches $A_0(m)\rightarrow 1/3$ in the limit $m \gg m_{h^0}$.

With these ingredients we construct the $\Delta \chi^2$ function. As $r_{gg}$ and $r_{\gamma\gamma}$ are correlated, we define a one parameter best-fit region in the physical stop parameter space at 2$\sigma$ C.L. by demanding $\Delta \chi^2 < 4$.

\subsection{Electroweak precision data}
Regarding the precision electroweak data, the largest effect of the stops and sbottoms is to provide a new one-loop contribution to the $\rho$ parameter~\cite{rho-susy}: 
\begin{eqnarray}
\Delta \rho & = &  \frac{3\,G_F}{8 \sqrt{2}\, \pi^2}
\bigg[   - s^2_t c^2_t \,F_0(m_{\tilde t_1}^2 , m_{\tilde t_2}^2)   \+ c^2_t \,F_0(m_{\tilde t_1}^2 , m_{\tilde b_L}^2) \+ s^2_t \,F_0(m_{\tilde t_2}^2 , m_{\tilde b_L}^2) \bigg], ~~~~~~~
\end{eqnarray}

where $G_F^{-1} = 2\sqrt{2}v^2$ and we have defined the function
\begin{equation}
F_0(x,y) = x+y -\frac{2 \,x \,y}{x-y}\log\frac{x}{y}.
\end{equation}
We have used the left-handed sbottom mass, $m_{\widetilde b_L}$, in \Eq{eq:sbottom} with the zero sbottom mixing. 
We apply the constraint from Ref.~\cite{Barger:2012hr}:
\begin{equation}
\Delta \rho = (4.2 \pm 2.7)  \times 10^{-4}.
\end{equation}

\subsection{Vacuum stability}  \bigskip

In the field space of $\{ \, H_u \approx h, \, {\widetilde t_L}, \, {\widetilde t_R} \, \}$, we study the (meta)stability conditions for our electroweak vacuum characterized by $\langle h \rangle/\sqrt{2} = v = 174$ GeV and vanishing stop vevs~\cite{vacuum-susy,Camargo-Molina:2014pwa}.
The tree-level potential that we use is (in terms of real scalar fields with $1/\sqrt{2}$ normalizations factored out)

\bea
V^0 &=& \frac{1}{2} m_H^2 h^2 \+ \frac{1}{2} m_Q^2 {\widetilde t_L}^2 \+ \frac{1}{2} m_U^2 {\widetilde t_R}^2 \+ \frac{y_t^2}{4} \Big( ({\widetilde t_L} {\widetilde t_R})^2 \+ (h {\widetilde t_L})^2 \+ ( h {\widetilde t_R})^2 \Big) \nonumber\\
&+&  \frac{g^{\prime 2}}{32} \Big( h^2 + \frac{1}{3} {\widetilde t_L}^2 - \frac{4}{3} {\widetilde t_R}^2 \Big)^2 \+ \frac{g^2}{32} \Big( h^2 \- {\widetilde t_L}^2 \Big)^2 \+ \frac{g_S^2}{24} \Big( {\widetilde t_L}^2 \- {\widetilde t_R}^2 \Big)^2 \\
&+& \frac{y_t X_t}{\sqrt{2}} \, h {\widetilde t_L}{\widetilde t_R} \+ h.c. \+ \frac{\delta \lambda}{4} h^4. \nonumber
\eea
We have assumed that the effects of $H_d$ are negligible, appropriate in the large $t_\beta$ regime, and have substituted $s_\beta \to 1$, $c_\beta \to 0$, and $H_u \to h/\sqrt{2}$.
To have a proper electroweak vacuum and the measured Higgs mass, we set $m_H^2 = -\frac{1}{2} m_h^2$, $m_h =125$ GeV, and $\delta \lambda = \frac{m_h^2}{2 v^2} - \frac{g^{\prime 2}}{8} - \frac{g^2}{8}$. 
In particular, we assume the presence of some physics, either radiative stop corrections in the MSSM or physics beyond the MSSM, that generates an appropriate correction to the Higgs quartic coupling, $\delta \lambda$. 
To arrive at the above form of the $g_S$ terms, we assume that the stops are aligned or anti-aligned in the $SU(3)_C$ color space.

The negative $X_t$\,-\,linear term can induce new minima when all three scalar fields obtain vevs. We numerically scan the field space to determine all local minima, and we use the \texttt{CosmoTransitions} program~\cite{Wainwright:2011kj} to calculate the tunnelling rate between our electroweak minimum and any deeper charge-color breaking minima. The classical action for this tunneling, $S_E \geq 400$ is required for the metastability of our vacuum.
We find that the parameter space with large values of the Higgs-stop-stop coupling, $\yhtt$, is constrained from the requirement of vacuum (meta)stability.

Our vacuum stability bounds rely on the simplifying assumptions discussed above, and it is possible that stronger constraints can be obtained if some of these assumptions are relaxed (e.g. $H_d$ vevs or misaligned stop vevs), or when one accounts for one-loop corrections to the potential and thermal effects (for a useful comparative study, see Ref.~\cite{Camargo-Molina:2014pwa}). Nevertheless we believe our bounds give a qualitatively correct representation of the current limits, and we observe that for light stops the vacuum stability limits are generally weaker than those coming from Higgs couplings.

When the left-handed stop is light, we also check whether the mass of the accompanying left-handed sbottom is positive (and evading LEP bounds). For the given stop parameters and heavy enough right-handed sbottoms, the left-handed sbottom is the heaviest when the sbottom mixing is zero (the equality below holds in this case)
\beq
m_{\widetilde b_L}^2 \, \leq \, \left( m_{\tilde t_1}^2 c_t^2 + m_{\tilde t_2}^2 s_t^2 \right) - m_t^2 - c_{2\beta} \, m_W^2.
\label{eq:sbottom} \eeq
Throughout in this work, we use this left-handed sbottom mass with the equality sign by assuming the zero sbottom mixing.

\section{Stop 3-body decay: $m_t + m_{\chi^0} > m_{\tilde t_1}  > m_W+ m_b + m_{\chi^0}$}
\label{app:decay}

The partial decay width for $\tilde t \rightarrow t^* \chi^0 \rightarrow W b \chi^0$ can be written as
\begin{equation}
\Gamma_{\tilde t_1 \rightarrow W b \chi^0} = \frac{1}{\pi} \int_{(m_W+m_b)^2}^{(m_{\tilde t_1} - m_{\chi^0})^2}   ds
 \frac{\sqrt{s}}{(s-m_t^2)^2 + m_t^2 \Gamma_t^2 } \, \Gamma_{\tilde t_1 \rightarrow t^* \chi^0}(s) \, \Gamma_{t^* \rightarrow Wb}(s),
 \end{equation}
where we have defined the ``off-shell" partial widths:
\begin{eqnarray}
\Gamma_{\tilde t_1 \rightarrow t^* \chi^0  }(s) & = & 
\frac{N_c m_{\tilde t_1}}{ 48 \pi}  \lambda^{1/2} \! \left(1,\frac{s}{m_{\tilde t_1}^2},\frac{m_{\chi^0}^2}{m_{\tilde t_1}^2}\right) 
 \left[ \left(g_L^2 \frac{m_t^2}{s}\! + g_R^2 \right) \left(\! 1 \! - \! \frac{s}{m_{\tilde t_1}^2} \! - \! \frac{m_{\chi^0}^2}{m_{\tilde t_1}^2}  \right) \! - \! 4 g_L g_R   \frac{m_t m_{\chi^0}}{m_{\tilde t_1}^2} \right] \nonumber \\
\Gamma_{t^*\rightarrow W b}(s)& =& \! \frac{N_c g^2 |V_{tb}|^2 }{192 \pi}\frac{s^{3/2}}{m_W^2}  \lambda^{1/2}\left( 1 , \frac{m_W^2}{s}  ,  \frac{m_b^2}{s}\right)
 \left[ \lambda\left(\! 1,\frac{m_W^2}{s},\frac{m_b^2}{s}\right)\! + \! 3\frac{m_W^2}{s} \left(\!1\!- \! \frac{m_W^2}{s}\! + \! \frac{m_b^2}{s}\right) \right], \nonumber \\
 &&
 \label{eq:3body}
 \end{eqnarray}
Here, we have defined the couplings,
\begin{eqnarray} 
g_L & = &  -\frac{2 \sqrt{2}}{3} \, g' \, N_{11}^* \, s_t + y_t \, N_{14}^* \, c_t, \nonumber  \\
g_R & = & \frac{1}{3\sqrt{2}} \, g' \, N_{11} \, c_t + \frac{1}{\sqrt{2}} \,g  \,N_{12} \, c_t + y_t \, N_{14} \, s_t,
 \end{eqnarray}
with $N_{ij}$ the neutralino mixing matrix elements, and the kinematic function $\lambda(a,b,c) = a^2+b^2+c^2-2 a b-2 a c-2 b c$. Notice that the three-body partial decay width reduces to the two-body one in the appropriate kinematical regime through the use of the narrow-width approximation.

\section{Limit Extrapolation to Future Searches}
\label{app:extrap}

To estimate the future prospects for probing stoponium at the LHC we extrapolate the current {\it expected} limits of the diboson resonance searches to $\sqrt{s}=14$ TeV with integrated luminosities ${\cal L}$ of 30/fb and 3/ab. To accomplish this, we employ several simplifying assumptions. 
For the uncertainties we consider two extreme cases: (1) statistical uncertainties dominate, and (2) systematic uncertainties dominate and improve in proportion to $\sqrt{{\cal L}}$. The results in the main text of Section~\ref{sec:limits} are obtained with the former assumption, but we will offer a comparison between the two cases below. 

\medskip

\underline{Statistical uncertainty dominant:}~ At a specified collision energy and integrated luminosity denoted collectively by a superscript $i$, we have 
\beq
\frac{\sigma_{\rm bound}^i}{\sigma_B^i} \propto \frac{1}{\sqrt{N_B^i}} \= \frac{1}{\sqrt{\hat{\sigma}_B {\cal P}_{q\bar{q}}^i {\cal L}^i \epsilon^i }},
\label{eq:stat}
\eeq
where $\sigma_{\rm bound}^i$ is the upper bound on the signal cross-section, $\sigma_B^i$  is the background cross-section, and $N_B^i$ is the number of background events after all cuts. The latter two quantities are related as $N_B^i = \sigma_B^i {\cal L}^i \epsilon^i = \hat{\sigma}_B {\cal P}_{q\bar{q}}^i {\cal L}^i \epsilon^i$, where $\hat{\sigma}_B$ is the partonic background cross-section (which does not depend on $i$), ${\cal L}^i$ is the integrated luminosity, $\epsilon^i$ is the selection efficiency, and 
 ${\cal P}_{ab}^i$ is the parton luminosity for initial partons $ab$, defined as 
\beq
{\cal P}^i_{ab} \, \equiv \, \int_{\tau^i}^1 dx \, \frac{\tau^i}{x} f_a(x, Q^2) f_b(\tau^i/x,Q^2), \qquad \tau^i \, \equiv \, \frac{ m_{\eta_{\tilde t}}^2 }{ s^i }.
\label{eq:Pab}
\eeq 
In Eq.~(\ref{eq:stat}) we have taken ${\cal P}_{q\bar q}^i $ since the dominant backgrounds in the most sensitive channels, $\gamma\gamma$ and $ZZ$, arise from $q\bar q$ initial states. In our numerics we take the up quark parton luminosity for simplicity, as it will dominate in $pp$ collisions. 

We furthermore assume a constant efficiency, $\epsilon^i=\epsilon$, which is reasonable since the signal-to-background ratio near a resonance is largely determined by the resonance mass. This is similar to gluino 
pair searches~\cite{Jung:2013zya} and perhaps more generally useful~\cite{colliderreach}. 

Taking ratios, we can extrapolate the current bounds to those for other collision energies and luminosities in a simple way in terms of parton luminosity ratio and luminosity ratio
\beq
\frac{\sigma_{\rm bound}^i}{\sigma_{\rm bound}^j} \= \frac{\sigma_B^i / \sqrt{N_B^i}}{ \sigma_B^j / \sqrt{N_B^j} } \=  \frac{ \sqrt{ \hat{\sigma}_B {\cal P}_{q\bar{q}}^i / {\cal L}^i } }{ \sqrt{ \hat{\sigma}_B {\cal P}_{q\bar{q}}^j / {\cal L}^j } } \= \sqrt{ \frac{ {\cal P}_{q\bar{q}}^i }{ {\cal P}_{q\bar{q}}^j } } \sqrt{ \frac{{\cal L}^j}{ {\cal L}^i } }.
\label{eq:statextra}\eeq

\medskip
\underline{Systematics dominant improving with 1/$\sqrt{{\cal L}}$:}~ We assume that systematic uncertainties improve with 1/$\sqrt{{\cal L}}$
\beq
 \frac{\sigma_{\rm bound}^i}{\sigma_B^i} \propto  \frac{1}{\sqrt{{\cal L}^i}},
\eeq
Again taking ratios, we are able to extrapolate the current limits:
\beq
\frac{\sigma_{\rm bound}^i}{\sigma_{\rm bound}^j} \= \frac{ \sigma_B^i / \sqrt{{\cal L}^i} }{  \sigma_B^j / \sqrt{{\cal L}^j} } \= \frac{ {\cal P}_{q\bar{q}}^i}{ {\cal P}_{q\bar{q}}^j} \sqrt{ \frac{ {\cal L}^j }{ {\cal L}^i } }.
\label{eq:sysextra}\eeq

\medskip

Comparing \Eq{eq:statextra} and \Eq{eq:sysextra}, we see that if errors are dominantly from systematics, one obtains a weaker cross-section bound by a factor $\sqrt{ {\cal P}^i_{q\bar{q}} / {\cal P}^j_{q\bar{q}}}$. Numerically, this equals to 1.45 and 1.73 for $m_{\widetilde t_1} =$ 160, 400 GeV at 14 TeV collision for $q = u$. 
These results roughly show how sensitively the future bounds may depend on the assumptions on errors.

\section{Resonance Searches} 
\label{app:searches}

In Table~\ref{tab:dataused}, we collect the latest experimental results on diboson resonance searches. The searches shown in bold give the strongest constraints in each diboson channel and are used to derive limits on stoponium. The CMS and ATLAS $\gamma \gamma$ limits are similar but constrain complementary mass ranges due to different statistical fluctuations, and we therefore show both results. The CMS $Z \gamma$ is found to be somewhat stronger than that of ATLAS, but due to the importance of this channel, we show both results and encourage more careful analysis. 
All searches use 8 TeV datasets except for the ones with $\sim$24/fb and $\sim$10/fb, in which case the combined 7+8 TeV dataset is used. The CMS $ZZ \to 4\ell$ result is stronger than the ATLAS one partly because CMS presents combined results of all production channels and uses the 7+8 TeV dataset. The $WW \to \ell \nu 2j$ result has similar sensitivity to that of the $WW \to 2 \ell 2 \nu$, but we show only the latter result for simplicity.
The searches listed in the last four lines give weaker constraints. Lastly, ATLAS $\gamma \gamma$ and $Z\gamma$ present limits on the fiducial cross-section, so we assume and unfold a constant fiducial efficiency for each channel as mentioned in the relevant part of text.

\begin{table}[t] \centering
\begin{tabular}{ c |  l l   |  l l }
\hline \hline
channel & \multicolumn{2}{c|}{CMS} & \multicolumn{2}{c}{ATLAS} \\
\hline \hline
$\gamma \gamma$ &  {\bf 19.7/fb} & {\bf HIG-14-006}~\cite{CMS:2014onr} &  {\bf 20.3/fb} & {\bf 1407.6583}~\cite{Aad:2014ioa}  \\
$Z \gamma \to \ell^+ \ell^- \gamma$ & {\bf 19.7/fb} & {\bf HIG-14-031}~\cite{CMS:2015lza} & {\bf 20.3/fb} & {\bf 1407.8150}~\cite{Aad:2014fha} \\
$ZZ \to 4\ell$ & {\bf 24.8/fb} & {\bf 1312.5353}~\cite{Chatrchyan:2013mxa} &  20.7/fb & CONF-2013-013~\cite{ATLAS:2013nma} \\
$WW \to \ell \nu 2j $ & {\bf 24.3/fb} & {\bf HIG-13-027}~\cite{CMS:2012bea} &   &  -- \\
$WW \to 2\ell 2\nu $ & {\bf 24.3/fb} & {\bf 1312.1129}~\cite{Chatrchyan:2013iaa} & 20.7/fb & CONF-2013-067~\cite{TheATLAScollaboration:2013zha}\\
$hh \to b \bar{b} \gamma \gamma$ & 19.7/fb & HIG-13-032~\cite{CMS:2014ipa} &  {\bf 20.0/fb} & {\bf 1406.5053}~\cite{Aad:2014yja} \\
\hline
$ZZ \to 2\ell 2j $ & 10.4/fb & 1304.0213~\cite{Chatrchyan:2013yoa} & 20.0/fb & CONF-2014-039~\cite{ATLAS:llqq} \\
$ZZ \to 2\ell 2\nu$ & 10.4/fb & 1304.0213~\cite{Chatrchyan:2013yoa}  & & -- \\
$hh \to 4b$ & 17.9/fb & 1503.04114~\cite{ Khachatryan:2015yea} & 19.5/fb & CONF-2014-005~\cite{ ATLAS:bbbb}  \\
$hh \to WWbb$ & & -- & 20.3/fb & 1312.1956~\cite{ Aad:2013dza} \\
\hline \hline
\end{tabular}
\caption{The latest data for heavy resonance searches used in this paper. Those in bold are used to derive limits on stoponium. Other results give weaker constraints.}
\label{tab:dataused}\end{table}


\end{document}